\documentclass[sigconf]{acmart}

\AtBeginDocument{%
  \providecommand\BibTeX{{%
    \normalfont B\kern-0.5em{\scshape i\kern-0.25em b}\kern-0.8em\TeX}}}

\setcopyright{acmcopyright}
\copyrightyear{2018}
\acmYear{2018}
\acmDOI{XXXXXXX.XXXXXXX}

\acmConference[Conference acronym 'XX]{Make sure to enter the correct
  conference title from your rights confirmation emai}{June 03--05,
  2018}{Woodstock, NY}
\acmPrice{15.00}
\acmISBN{978-1-4503-XXXX-X/18/06}

\acmSubmissionID{7202}

\usepackage{subfigure}
\usepackage{float}
\usepackage[ruled]{algorithm2e}
\usepackage{hyperref}
\usepackage{multirow}
\usepackage{adjustbox}

\usepackage[switch]{lineno}

\begin{document}
\title{Enhancing Recommender Systems: A Strategy to Mitigate False Negative Impact}

\author{Kexin Shi}
\affiliation{%
  \institution{the Hong Kong University of Science\\ and Technology}
  \country{}}
\email{kshiaf@connect.ust.hk}
\author{Yun Zhang}
\affiliation{%
  \institution{the Hong Kong University of Science\\ and Technology}
  \country{}}
\email{yzhangjy@connect.ust.hk}
\author{Wenjia Wang}
\affiliation{%
  \institution{Hong Kong University of Science\\ and Technology (Guangzhou)}
  \country{}}
\email{wenjiawang@ust.hk}
\author{Bingyi Jing}
\affiliation{%
  \institution{Southern University of Science\\ and Technology}
  \country{}}
\email{jingby@sustech.edu.cn}


\begin{abstract}
In implicit collaborative filtering (CF) task of recommender systems, recent works mainly focus on model structure design with promising techniques like graph neural networks (GNNs). Effective and efficient negative sampling methods that suit these models, however, remain underdeveloped. One challenge is that existing hard negative samplers tend to suffer from severer over-fitting in model training. In this work, we first study the reason behind the over-fitting, and illustrate it with the incorrect selection of false negative instances with the support of experiments. In addition, we empirically observe a counter-intuitive phenomenon, that is, polluting hard negative samples’ embeddings with a quite large proportional of positive samples’ embeddings will lead to remarkable performance gains for prediction accuracy. On top of this finding, we present a novel negative sampling strategy, \emph{i.e.}, positive-dominated negative synthesizing (PDNS). Moreover, we provide theoretical analysis and derive a simple equivalent algorithm of PDNS, where only a soft factor is added in the loss function. Comprehensive experiments on three real-world datasets demonstrate the superiority of our proposed method in terms of both effectiveness and robustness. The source code and data will be released upon the paper's acceptance.
\end{abstract}





\maketitle

\section{Introduction}
\label{sec:intr}
\begin{figure}[ht]
\centering
\includegraphics[width=0.45\textwidth]{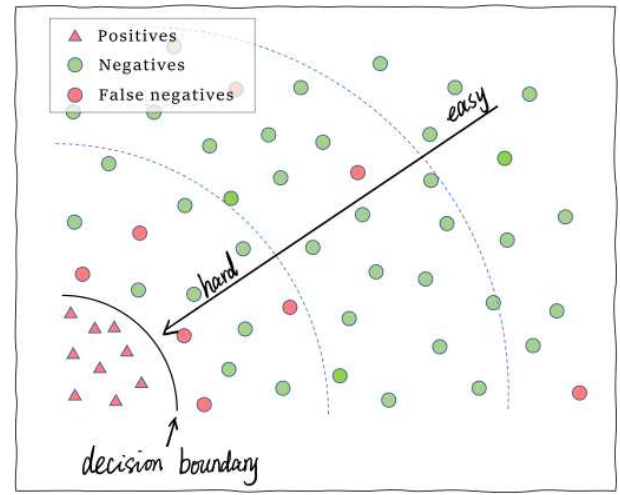}
\caption{In hard negative sampling of recommender systems, the harder a negative becomes, the more likely it is a false one, the more difficult to distinguish hard negatives from false negatives.}
\label{fig.falseneg}
\end{figure}

Recently, recommender systems have been widely deployed in various scenarios to avoid information overload, \emph{e.g.}, the personalized recommendations of online music, movies or commodities~\citep{moscato2020emotional,chen2019top,sun2018conversational,Zhang2017DeepLB,Tang2018AdversarialTT,Pan2020LearningSR}. Implicit collaborative filtering (CF) task in recommender systems is to predict the preferences of users according to their implicit feedback~\citep{Hu2008CollaborativeFF,liang2018variational,Najafabadi2016ASL,zheng2016neural,lee2010collaborative,DBLP:conf/kdd/Koren08,DBLP:conf/sigir/0001DWLZ020,DBLP:conf/kdd/HuangDDYFW021}. Compared to conventional methods in implicit CF, such as matrix factorization~\citep{DBLP:conf/kdd/Koren08,DBLP:conf/nips/SalakhutdinovM07,DBLP:journals/computer/KorenBV09,DBLP:journals/corr/DziugaiteR15,lee2000algorithms,mnih2007probabilistic}, graph neural networks (GNNs)~\citep{DBLP:conf/kdd/Wang00LC19, DBLP:journals/corr/BergKW17,DBLP:conf/sigir/Wang0WFC19, DBLP:conf/sigir/0001DWLZ020,DBLP:conf/kdd/HuangDDYFW021} have emerged as the most promising direction in accurate prediction of user preferences. Recently proposed GNN-based models GCMC~\citep{DBLP:journals/corr/BergKW17}, LightGCN~\citep{DBLP:conf/sigir/0001DWLZ020}, and KGAT~\citep{DBLP:conf/kdd/Wang00LC19} significantly improve the model prediction accuracy by treating interactions between users and items as a graph and mining more fine-grained relations among nodes. While most prior efforts focus on recommendation model structure design, another critical component of recommender training, \emph{i.e.}, negative sampling, receives less attention.

Negative sampling is an essential component in implicit CF task. Specifically, in most cases, we can collect users’ implicit feedback, \emph{e.g.}, product purchase or video watching history, and regard these observed user-item interaction pairs as positive targets, since the historical interactions may indicate users’ potential interests. For the remaining large number of unobserved interactions, we need to conduct negative sampling to pick possible negative user-item pairs, and try to avoid false negatives that users may like but have not yet interacted with. Then, the model is optimized by giving higher rank scores to positive pairs against negatives.

In previous works, there are several branches towards negative sampling. The simplest way is to sample unobserved interactions according to a manually designed fixed distribution. For example, uniform distribution is widely adopted in recommender systems~\citep{DBLP:conf/kdd/Wang00LC19, DBLP:conf/sigir/0001DWLZ020,DBLP:conf/kdd/HuangDDYFW021, DBLP:conf/sigir/Wang0WFC19}. Subsequently, to accelerate the convergence and enhance the effectiveness of recommendation models, hard negative sampling has been devised. This technique aims to derive high-quality and informative negative samples from extensive unobserved interaction data. For example, GAN-based samplers~\citep{DBLP:conf/sigir/WangYZGXWZZ17,DBLP:conf/www/ParkC19,DBLP:conf/ijcai/DingQ00J19,wang2018incorporating} leverage generative adversarial nets (GANs) to iteratively generate hard negative instances to fool the discriminator, and dynamic hard samplers~\citep{zhang2013optimizing,shi2023theories,DBLP:conf/nips/DingQY0J20, DBLP:conf/wsdm/RendleF14, DBLP:conf/kdd/HuangSSXZPPOY20,DBLP:conf/kdd/HuangDDYFW021} adaptively select negatives with high matching scores with users to pose challenges to the learning process of the recommendation model. Above hard samplers generally achieve performance gains over the fixed-distribution sampling approaches. Nevertheless, a collection of hard sampling techniques~\citep{zhang2013optimizing,DBLP:conf/wsdm/RendleF14,DBLP:conf/kdd/HuangDDYFW021,DBLP:conf/sigir/WangYZGXWZZ17,DBLP:conf/www/ParkC19} are demonstrated less robust, that is, they tend to suffer from severe over-fitting at the late stage of training process. Their performances soar in the beginning, but drop dramatically soon after. Therefore, a natural question arises: \emph{What is the reason behind the over-fitting?}
\\\\
\textbf{Contributions.}
In this paper, we give an answer to the above question. First, we investigate why over-fitting occurs when leveraging hard negative sampling strategies, by controlling the hardness level of negative samples and examining the severity of the over-fitting. We empirically find that increasing the hardness level will exacerbate its severity, on top of which, we attribute the over-fitting to the incorrect selection of false negatives during hard negative sampling. Moreover, we support this claim through simulation experiments conducted on two synthetic datasets.

This work presents a hard negative synthesizing strategy PDNS to avoid false negatives in negative sampling, and thus alleviate over-fitting in recommender systems. In PDNS, we synthesize hard negative instances by incorporating positive embeddings into negative embeddings, and more importantly, the synthetic negatives are dominated by the positive information rather than negative information. Comprehensive experiments demonstrate that PDNS not only largely mitigates the over-fitting, but also obtains a significant performance gain in terms of effectiveness. 

Meanwhile, we offer theoretical analysis on the mechanism of PDNS, suggesting that PDNS has a capacity to resist assigning very large gradient magnitudes to the hardest negatives among all hard negative samples. This behavior implicitly reduces the risk of automatically raising samples’ hardness level when updating the model, and thus making PDNS more robust to false negatives. Based on this analysis, we present a simple equivalent algorithm of PDNS, where solely the loss objective is slightly modified with a soft factor. Extensive experiments exhibit that PDNS not only suits GNN-based methods, but also can be applied to other types of models, such as MF.

To summarize, the contributions of our work are as follows:

\begin{itemize}
\item We suggest that the incorrect selection of false negatives contributes to the over-fitting that occurs in implicit CF when adopting hard negative sampling, and we verify it through simulation experiments.

\item We propose a general hard negative mining strategy PDNS for recommender systems, and theoretically reveal that PDNS is robust to false negative instances during hard negative sampling. 

\item We demonstrate the advantages of PDNS over a set of state-of-the-art negative sampling approaches in terms of robustness and effectiveness by experimenting on three real-world datasets.
\end{itemize}

\section{Preliminaries and problem}
In this section, we first introduce Bayesian personalized ranking (BPR), and then give an overview of hard negative sampling approaches. Finally, as one of contributions, we empirically study the over-fitting problem that exists in hard negative mining cases.

\subsection{Formulation}

\begin{figure*}[htp]
\centering
\subfigure[Over-fitting, Taobao]{
\label{fig.overfit.taobao}
\includegraphics[width=0.24\textwidth]{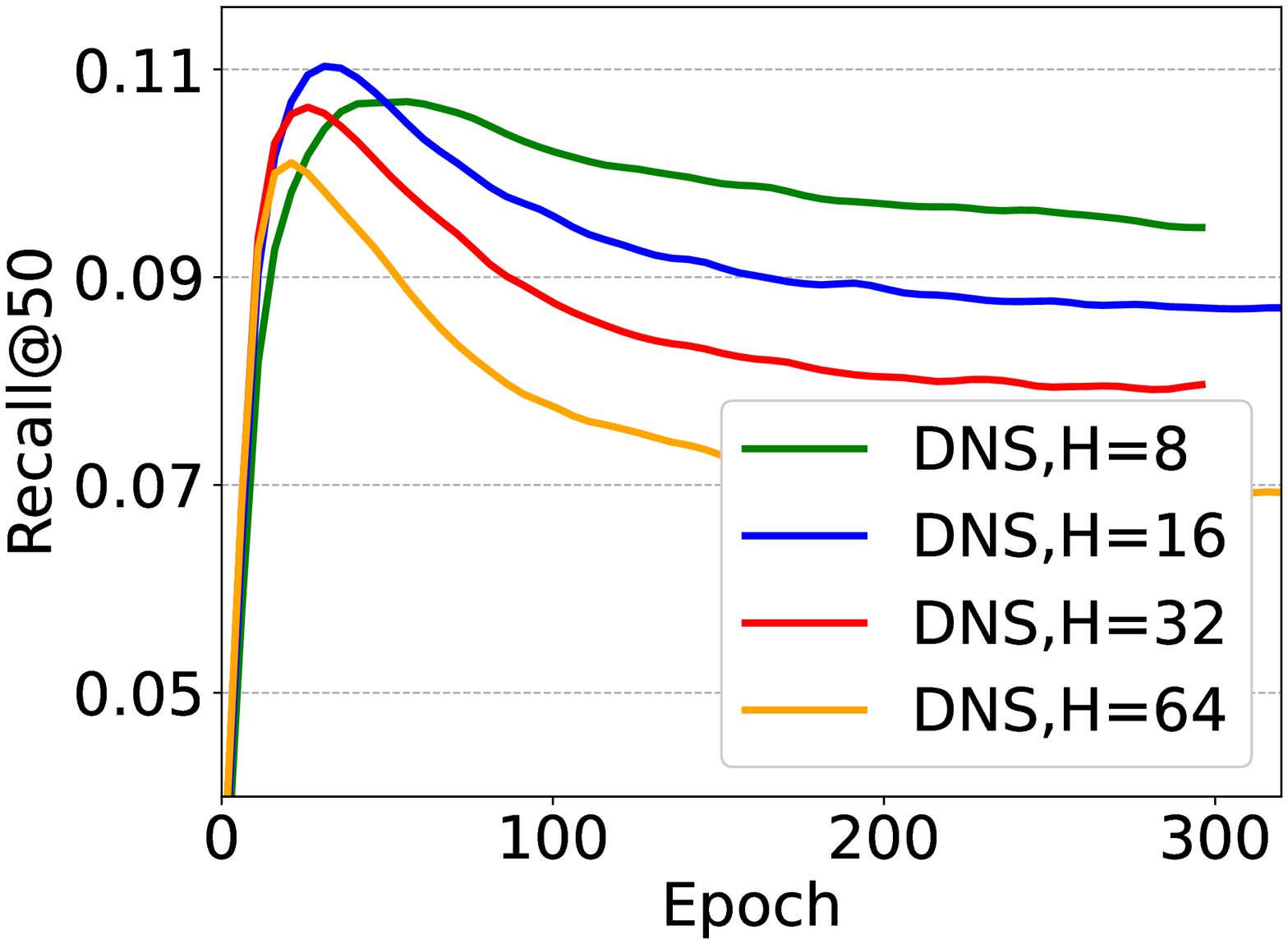}}
\subfigure[Over-fitting, Tmall]{
\label{fig.overfit.tmall}
\includegraphics[width=0.24\textwidth]{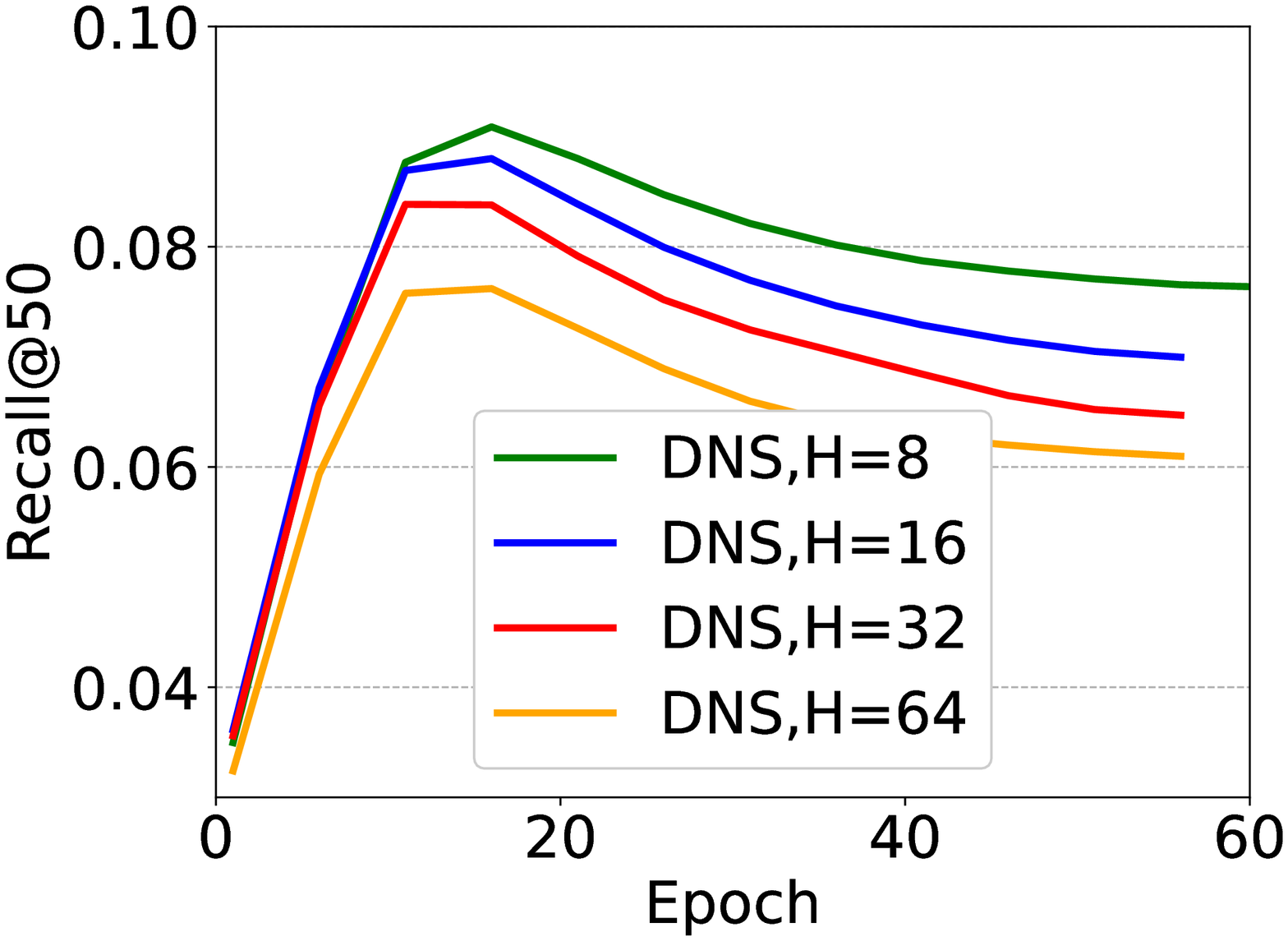}}
\subfigure[Positive mixing, Taobao]{
\label{fig.pm.taobao}
\includegraphics[width=0.24\textwidth]{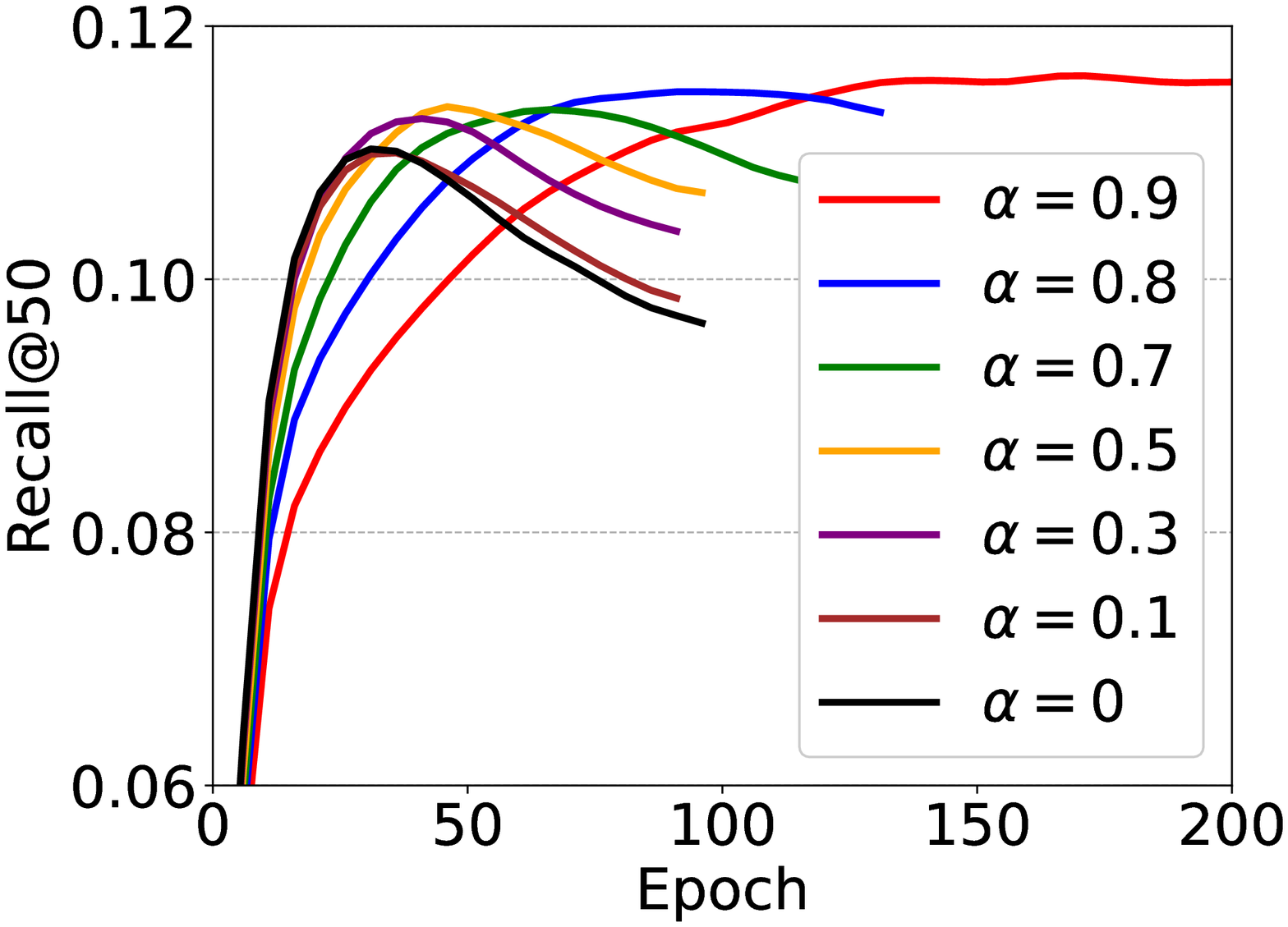}}
\subfigure[Positive mixing, Tmall]{
\label{fig.pm.tmall}
\includegraphics[width=0.24\textwidth]{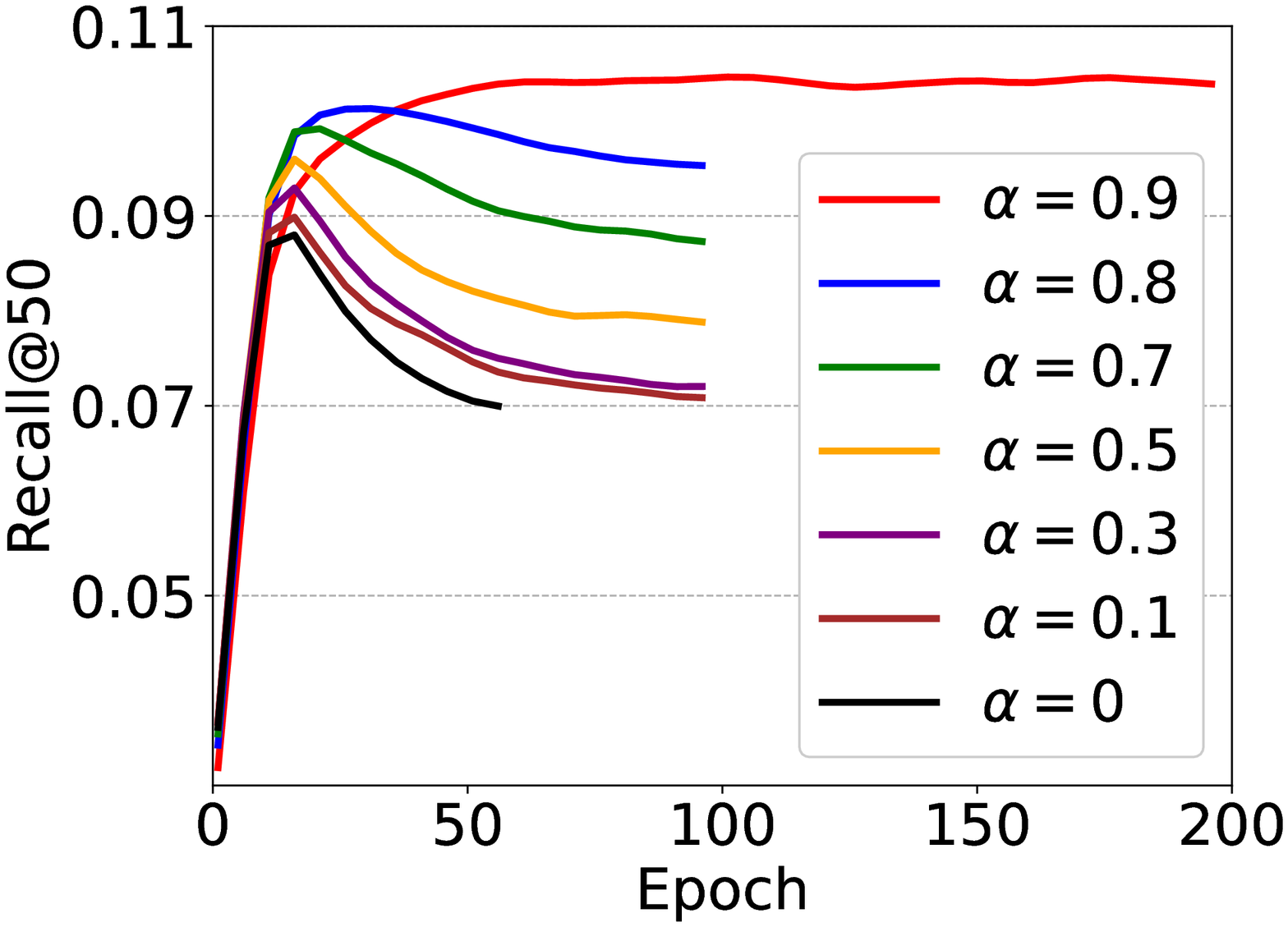}}
\caption{(a)-(b) Over-fitting problem of DNS on two datasets. (c)-(d) Effect of positive mixing in DNS.}
\label{fig.overfit and positive mixing}
\end{figure*}

We define $\mathcal{S}=\left\{(u,i)\right\}$ as the training set, containing all observed interactions between each user $u \in \mathcal{U}$ and each item $i \in \mathcal{I}$, (\emph{i.e.} positive user-item pair $(u,i)$). Note that such positive user-item pairs indicate user $u$ may have a preference on item $i$. The goal of recommendation is to predict users’ preferences and generate a ranked list of items for each user $u$ for personalized recommendation. To achieve this, a common strategy is to sample a number of unobserved pairs $\left\{(u,j)\mid j \in \mathcal{I} \setminus \mathcal{I}_{u} \right\}$ for user $u$ by a negative sampling function $f_{\mathrm{neg}}$, and treat them as negative samples, where $\mathcal{I}_{u}$ is the set of items that have interactions with user $u$. Then, a base recommendation model such as LightGCN~\citep{DBLP:conf/sigir/0001DWLZ020} is optimized by giving a higher score towards $(u,i)$ than $(u,j)$. Following this idea, a widely-adopted loss objective in recommender systems, BPR loss~\citep{DBLP:conf/uai/RendleFGS09}, is designed as follows:
\begin{equation}
    \mathcal{L}_{\mathrm{BPR}}=-\sum_{\substack{(u,i)\in \mathcal{S}\\j\sim f_{\mathrm{neg}}}} \ln \sigma \left(\hat{y}_{ui}-\hat{y}_{uj}\right),\\
\label{bprloss}
\end{equation}
\begin{equation}
    \text{ with }\ \hat{y}_{ui}=\mathbf{e}_u\cdot \mathbf{e}_i, \textbf{  } \hat{y}_{uj}=\mathbf{e}_u\cdot \mathbf{e}_j,
\end{equation}
where $\mathbf{e}_{*}$ is the embedding of $*$ generated by the recommendation model, and we apply dot product function to calculate the user-item pair scores. Here, $\sigma (\cdot)$ is a sigmoid function. In this work, we use above BPR loss for model optimization. 

This paper focuses on the negative sampling function $f_{\mathrm{neg}}$. We derive user and item embeddings by leveraging a state-of-the-art GNN-based model LightGCN due to its excellent performance over other baselines. However, the negative sampling method $f_{\mathrm{neg}}$ we propose can be generalized to other types of recommendation models like MF~\citep{DBLP:conf/kdd/Koren08} (see experiments in Section~\ref{sec:real_exp}).

\subsection{Hard negative sampling and over-fitting problem}

\label{sec:over-fitting}
As mentioned in Introduction, uniform sampling has been commonly deployed in recommender systems, which regards all unobserved items $\left\{j | j \in 
\mathcal{I} \setminus \mathcal{I}_u\right\}$ having the same probability to be negative for user $u$. But in order to improve the quality of negatives, many prior efforts work on picking more informative negatives (\emph{a.k.a.} hard negatives) by preferring those unobserved items with relatively high matching scores with the user. Including hard negatives in training can be beneficial in enhancing the discriminative power of the recommendation model, since the model is forced to learn more fine-grained features that effectively separate positives from negatives. However, in most works towards hard negative sampling, over-fitting issue is reported~\citep{DBLP:conf/kdd/HuangDDYFW021,DBLP:conf/sigir/WangYZGXWZZ17,DBLP:conf/wsdm/RendleF14,DBLP:conf/www/ParkC19}. As in Figure~\ref{fig.overfit.taobao} and \ref{fig.overfit.tmall}, the recommender performance drops sharply in the late phase of training process.
\\\\
\textbf{Over-fitting.} To further explore what causes the over-fitting in hard negative sampling of recommender systems, we conduct experiments with a representative hard negative sampling strategy DNS~\citep{DBLP:conf/wsdm/RendleF14}, which serves as a foundation approach that encapsulates the core principles of other hard mining techniques. Specifically, given a positive user-item pair $(u, i)$, we first form the negative candidate pool $\mathcal{C} = \left\{j_1, j_2 \dots, j_H \right\}$ of size $H$ by uniformly sampling items from non-interacted items $\left\{j | j \in 
\mathcal{I} \setminus \mathcal{I}_u\right\}$. Then, the hardest negative is selected as: 
\begin{equation}
j = \mathop{\arg\max}_{k \in \mathcal{C}} \mathbf{e}_u \cdot \mathbf{e}_k,
\label{step:DNS}
\end{equation}
where $\mathbf{e}_u$ and $\mathbf{e}_k$ are the embeddings of user $u$ and item $k$, respectively, obtained through a recommendation model, like LightGCN.

\label{state_H}
Note that when $H$ (the size of negative candidate pool) is larger, negative samples selected by Eq.\eqref{step:DNS} are generally harder. Conversely, when $H=1$, it degenerates to a uniform sampling. Thus, we also use $H$ to represent the hardness level of negatives. We conduct experiments on two real-world datasets Taobao and Tmall (details in Table~\ref{tab:data}) with $H$ varying in $\left\{8,16,32,64\right\}$, and draw their training curves in Figure~\ref{fig.overfit.taobao}-\ref{fig.overfit.tmall}. Some observations are as follows:
\begin{itemize}
\item {Over-fitting problem exists in all training curves with different negative hardness setting $H$. The performances of DNS soar in the beginning, then degrade dramatically soon after.}
\item {The harder the negatives selected, the severer the over-fitting. For example, the training curve with $H=64$ decreases faster than the one with $H=8$ on Taobao.}
\item {The best performance is achieved at an intermediate value ($H=16$ for Taobao and $H=8$ for Tmall). It is not the case that the recommender performs better along with the increase of the hardness level $H$.\\ }
\end{itemize}
\textbf{Explanation.} Here, we provide an explanation, that is, the reason why above three observations about over-fitting occur is that the hard negative sampler tends to incorrectly select plenty of false negatives for model optimization, especially in the late training phase. These false negatives will provide misleading information, and thus lead to performance degradation.

To be specific, false negatives are potential positive samples yet are selected as negatives. False negatives are generally highly scored by the recommender, thus it is difficult to distinguish them and real hard negatives. In addition, \citet{DBLP:conf/nips/DingQY0J20} shows that the harder the negative, the more likely it is a false one (as shown in Figure~\ref{fig.falseneg}). Therefore, we illustrate above three observations as follows:

\begin{itemize}
\item At the beginning of model training, negatives selected in negative sampling are relatively random due to the limited accuracy of the recommender, thus it is less likely to encounter false negatives. The performance improves rapidly. With the recommender becoming increasingly accurate, however, the selected negatives get harder, leading to a rapid decline of performance due to their large possibility to be the false ones. 

\item As for the last two observations, when the hardness level $H$ increases, the model updates can benefit from real harder (more informative) negative instances, but can also be misled by those incorrectly selected false negatives. Hence, best performance is achieved in the trade-off between real hard negatives’ informativeness and false negatives’ corruption.
\end{itemize}

To summarize, for hard negative mining, the lack of robustness to false negative instances contributes to the over-fitting in model training of implicit CF. We also conduct simulation experiments in Section~\ref{sec:syn_experiment} for verification.

\section{method: Positive-dominated negative synthesizing}
In this section, we first study the influence of positive mixing technique on recommender training, which inspires us to propose a positive-dominated negative synthesizing method (PDNS) to generate hard negatives. Next, we offer a theoretical analysis on the loss function, disclosing that PDNS is robust to false negative instances, and we provide a simple equivalent algorithm of PDNS.
\subsection{Positive mixing study}
 Positive mixing is a technique to generate synthetic harder negatives from existing negative samples by injecting positive information~\citep{DBLP:conf/nips/KalantidisSPWL20}. Specifically, for user $u$, positive item $i\in\mathcal{I}_{u}$ and negative item $j \in \mathcal{I} \setminus \mathcal{I}_u$, the embedding of synthetic negative instance from positive mixing, $\widetilde{\mathbf{e}}_{j^{’}}$, can be formulated as follows:
\begin{equation}
\widetilde{\mathbf{e}}_{j^{’}} = \alpha \mathbf{e}_i + (1-\alpha) \mathbf{e}_j,  \alpha \in P(\cdot)
\label{positivemixing}
\end{equation}
where $\alpha$ is the mixing coefficient that can be drawn from pre-defined distributions $P(\cdot)$, $\mathbf{e}_i$ and $\mathbf{e}_j$ are embeddings of items $i$ and $j$ respectively, obtained through the recommendation model.

In recent works, positive mixing strategy is generally coupled with other techniques, \emph{e.g.}, hardest mixing~\citep{DBLP:conf/nips/KalantidisSPWL20} and hop mixing~\citep{DBLP:conf/kdd/HuangDDYFW021} to achieve state-of-the-art performances. They claim that positive mixing works as it can generate harder ones compared to original negatives. In MixGCF~\citep{DBLP:conf/kdd/HuangDDYFW021}, the mixing coefficient $\alpha$ is randomly drawn from uniform distribution $\mathbf{U}(0,1)$, and in MoCHi~\citep{DBLP:conf/nips/KalantidisSPWL20}$, \alpha$ is randomly drawn from uniform distribution $\mathbf{U}(0,0.5)$ to ensure that the contribution of the positive is smaller than that of the negative. However, there is a lack of theoretical analysis on how the mixing coefficient $\alpha$ affects the recommender performance.

To explore the influence of the mixing coefficient $\alpha$, here, we set $\alpha$ as a fixed number and vary it in $\{0,0.1,0.2,\cdots,0.9\}$. Given item $j$ that is chosen by DNS as a hard negative for user $u$, the synthetic negative embedding from positive mixing, $\widetilde{\mathbf{e}}_{j^{’}}$, is derived by Eq.\eqref{positivemixing}. Note that when $\alpha =0$, there is no positive information injected, thus it degenerates to basic DNS. As shown in Figure~\ref{fig.pm.taobao}-\ref{fig.pm.tmall}, We observe that:

\begin{itemize}
    \item In terms of effectiveness, it achieves better performance with a large $\alpha$, and yields the best result when $\alpha=0.9$, which outperforms DNS (i.e. $\alpha=0$) by a large margin.
    \item In terms of robustness, adopting positive mixing with a large $\alpha$ can significantly mitigate over-fitting problem.
\end{itemize}

These two findings are counter-intuitive yet critical, which can be leveraged to enhance the model performance considerably. However, they are neglected by prior efforts that pay much attention on multi-technique combinations. 

\subsection{Positive-dominated negative synthesizing (PDNS)}


\label{sec:pdns}
Motivated by above findings, in order to enhance robustness and effectiveness of the recommender, we propose a novel hard negative sampling strategy PDNS. As an advanced version of DNS, PDNS is easy to implement but efficient, which consists of two steps, \emph{i.e.}, basic DNS selection and positive-dominated mixing.
\\\\
\textbf{Positive-dominated mixing.} With the observed positive item $i$ and the hard negative item $j$ selected by DNS (details are described in Section~\ref{sec:over-fitting}), we synthesize a new hard instance $j^{’}$ by injecting a large proportion of positive embedding $\mathbf{e}_i$ into $\mathbf{e}_j$, formulated as follows:
\begin{equation}
 \widetilde{\mathbf{e}}_{j^{’}} = \alpha \mathbf{e}_i + (1-\alpha) \mathbf{e}_j.
 \label{pdns}
\end{equation}
Notably, $\alpha$ here is a constant coefficient set normally larger than $0.7$ to ensure the domination of positive information. 
\\\\
\textbf{Loss function analysis.}
\label{sec:theo}
To further study the mechanism of PDNS, we take a closer look at two loss objectives \emph{w.r.t.} DNS and PDNS, separately. Note that in the following functions, leveraging DNS and PDNS as the negative sampling strategy is termed $f_{\mathrm{DNS}}$ and $f_{\mathrm{PDNS}}$ respectively.
\newline
\newline
By Eq.\eqref{bprloss}, BPR loss with DNS can be written as:
\begin{equation}
\begin{aligned}
    \mathcal{L}_{\mathrm{DNS}} &=-\sum_{\substack{(u,i)\in \mathcal{S}\\j\sim f_{\mathrm{DNS}}}} \ln \sigma \left(\mathbf{e}_u \cdot \mathbf{e}_i -   \mathbf{e}_u \cdot \mathbf{e}_j\right) \\
    &\triangleq -\sum_{\substack{(u,i)\in \mathcal{S}\\j\sim f_{\mathrm{DNS}}}} \ln \sigma \left(\hat{y}_{ui}-\hat{y}_{uj}\right).
\end{aligned}
\label{dnsbpr}
\end{equation}
\newline
\newline
Comparatively, BPR loss with PDNS can be written as:
\begin{equation}
\begin{aligned}
    \mathcal{L}_{\mathrm{PDNS}} &=-\sum_{\substack{(u,i)\in \mathcal{S}\\j^{\prime}\sim f_{\mathrm{PDNS}}}} \ln \sigma \left(\mathbf{e}_u \cdot \mathbf{e}_i -   \mathbf{e}_u \cdot \widetilde{\mathbf{e}}_{j^{\prime}}\right) \\
    &= -\sum_{\substack{(u,i)\in \mathcal{S}\\j\sim f_{\mathrm{DNS}}}} \ln \sigma \left(\mathbf{e}_u \cdot \mathbf{e}_i - \mathbf{e}_u \cdot \left(\alpha \mathbf{e}_i+(1-\alpha)\mathbf{e}_j\right)\right)\\
    &= -\sum_{\substack{(u,i)\in \mathcal{S}\\j\sim f_{\mathrm{DNS}}}} \ln \sigma \left((1-\alpha)(\hat{y}_{ui}-\hat{y}_{uj})\right).
\end{aligned}
\label{pdnsbpr}
\end{equation}
\newline
\newline
Gradient-based optimizers, such as Adam~\citep{kingma2014adam}, are generally used to minimize loss objectives. Thus, here, we investigate into the difference between gradients of $\mathcal{L}_{\mathrm{PDNS}}$ and $\mathcal{L}_{\mathrm{DNS}}$. The gradient of $\mathcal{L}_{\mathrm{DNS}}$ \emph{w.r.t} parameter $\theta \in \Theta$ of the recommendation model like LighGCN, is as follows:
\begin{equation}
    \frac{\partial \mathcal{L}_\mathrm{DNS}}{\partial \theta} = -\sum_{\substack{(u,i)\in \mathcal{S}\\j\sim f_{\mathrm{DNS}}}}\left(1-\sigma \left(\hat{y}_{ui}-\hat{y}_{uj}\right)\right)\frac{\partial \left(\hat{y}_{ui}-\hat{y}_{uj}\right)}{\partial \theta},
    \label{dnsgradient}
\end{equation}
where for each triplet $(u,i,j)$, we denote its multiplicative scalar $1-\sigma \left(\hat{y}_{ui}-\hat{y}_{uj}\right)$ as $\Delta_{(u,i,j)}^{\mathrm{DNS}}$. It reflects how much the recommendation model learns from the triplet $(u,i,j)$.
\newline
\newline
Comparatively, the gradient of $\mathcal{L}_{\mathrm{PDNS}}$ is:
\begin{equation}
    \frac{\partial \mathcal{L}_\mathrm{PDNS}}{\partial \theta} = -(1-\alpha)\sum_{\substack{(u,i)\in \mathcal{S}\\j\sim f_{\mathrm{DNS}}}}\left(1-\sigma \left((1-\alpha)\left(\hat{y}_{ui}-\hat{y}_{uj}\right)\right)\right)\frac{\partial \left(\hat{y}_{ui}-\hat{y}_{uj}\right)}{\partial \theta},
    \label{pdnsgradient}
\end{equation}
where $1-\sigma \left((1-\alpha)\left(\hat{y}_{ui}-\hat{y}_{uj}\right)\right)$ is defined as $\Delta_{(u,i,j)}^{\mathrm{PDNS}}$.
\newline
\newline

\indent Note in Eq.\eqref{pdnsgradient}, the first constant coefficient term $(1-\alpha)$ has no influence on model training with optimizer Adam used in our experiments. The only difference between gradients of $\mathcal{L}_\mathrm{PDNS}$ and $\mathcal{L}_\mathrm{DNS}$ is that the sigmoid function $\sigma (\cdot)$ is stretched flatter via multiplying by a small number $1-\alpha$, indicating that only stretching $\sigma(\cdot)$ in BPR loss can significantly improve the recommender performance regarding robustness and effectiveness.

To be specific, in DNS, the $\Delta_{(u,i,j)}^{\mathrm{DNS}}$ of different triplets $(u, i, j)$ differ greatly, some of which are near 0 if $\hat{y}_{ui}$ is correctly scored higher than $\hat{y}_{uj}$, but some of which are near 1 if the opposite happens. In the case that $\Delta_{(u,i,j)}^{\mathrm{DNS}}\approx 0$, the recommender learns almost nothing from $(u,i,j)$ triplet, leading to the waste of some selected hard negatives, and forcing model updates rely on the remaining harder ones. In fact, this behavior of BPR loss makes the parameter update based on negatives which are much harder than expectation. Hence, there is a higher likelihood of encountering false negatives (considering that harder negatives are more prone to being false ones), resulting in severe over-fitting. Meanwhile, the triplets $(u,i,j)$ with $\Delta_{(u,i,j)}^{\mathrm{DNS}}\approx 0$ are useless in training process, that makes the model less effective due to the loss information.

Comparatively, in PDNS, $\sigma(\cdot)$ is stretched to be flatter, in this way differences among $\Delta_{(u,i,j)}^{\mathrm{PDNS}}$s are reduced. The effectiveness of the recommendation model is enhanced, since more hard negatives selected from DNS can be involved in training process. Meantime, the algorithm is more robust, because involving more hard negatives in model updating rather than only focusing on the hardest ones can reduce the probability of false negative selection.

\subsection{Equivalent algorithm of PDNS}
\label{equi_alg}
\begin{algorithm}
\SetKwInOut{Input}{Input}
\caption{Equivalent algorithm of PDNS}
\label{alg}
\Input{Training set $\mathcal{S}=\left\{(u,i)\right\}$, recommendation model such as LightGCN with learnable parameter $\Theta$, the size of negative candidate pool $H$, the soft factor $\beta$.}
\For{$t=1,2,\cdots,T$}{
Sample a mini-batch $\mathcal{S}_{batch}\in \mathcal{S}$; 
Initialize loss $\mathcal{L}_{\mathrm{BPR}}^{\mathrm{soft}}$\;
\For{each $\left(u,i\right) \in \mathcal{S}_{batch}$}{
Get the embeddings of users and items through the recommendation model;

Form the negative candidate pool $\mathcal{C}$ by uniformly sampling $H$ items from $\left\{j\mid j\in \mathcal{I} \setminus \mathcal{I}_u\right\}$;

Get the hardest negative item $j$ from $\mathcal{C}$;

Update $\mathcal{L}_{\mathrm{BPR}}^{\mathrm{soft}}$ \eqref{soft_bprloss}.}
Update $\theta$ according to gradient \emph{w.r.t.} $\mathcal{L}_{\mathrm{BPR}}^{\mathrm{soft}}$ \eqref{soft_bprloss}.}
\end{algorithm}

Now, we present an equivalent algorithm of PDNS by simply modifying BPR loss. By comparing $\mathcal{L}_\mathrm{DNS}$ and $\mathcal{L}_\mathrm{PDNS}$, it is obvious that applying positive-dominated mixing has the same effect as scaling the sigmoid function in $\mathcal{L}_\mathrm{DNS}$. Therefore, we propose a modified BPR loss as follows:
\begin{equation}
\mathcal{L}_{\mathrm{BPR}}^{\mathrm{soft}}=-\sum_{\substack{(u,i)\in \mathcal{S}\\j\sim f_{\mathrm{neg}}}} \ln \sigma \left(\beta\left(\hat{y}_{ui}-\hat{y}_{uj}\right)\right).
\label{soft_bprloss}
\end{equation}    
Here, the loss function $\mathcal{L}_{\mathrm{BPR}}^{\mathrm{soft}}$ is a soft BPR loss, where $\beta$ is the soft factor, normally tuned within the range of $(0,0.3)$ for better performance. 

Thus, we develop PDNS into DNS equipped with soft BPR loss, as shown in Algorithm~\ref{alg}. It is an easy-implemented framework, solely adding a soft factor to BPR loss while maintaining DNS unchanged. In Section~\ref{sec:real_exp}, Algorithm~\ref{alg} is utilized for extensive experiments and analysis.
\\\\
\textbf{Complexity analysis.} As in Algorithm~\ref{alg}, PDNS has the same time complexity and model complexity with DNS, demonstrating one aspect of advantages over other relatively complicated hard negative mining techniques, like the time-consuming GAN-based samplers. The selection scheme we present has $O(HT)$ time complexity, where $H$ is the size of negative candidate pool, $T$ denotes the time cost of each score computation. The model complexity of PDNS is $O((|\mathcal{U}|+|\mathcal{I}|)F)$, where $F$ represents the embedding dimension of users and items. 

\section{EXPERIMENTS}
\label{sec:exp}
We first conduct simulation experiments on two synthetic datasets to verify the impact of false negative instances on recommenders’ over-fitting. We then evaluate the performance of PDNS on three real-world datasets, with LightGCN and MF as the underlying recommendation models.

\subsection{Experimental settings}

\begin{table*}[htp]
  \begin{center}
    \caption{Details of datasets.}
    \label{tab:data}
    \begin{adjustbox}{width=0.6\textwidth}
    \begin{tabular}{lrrrrrr}
    \hline &Dataset& User  & Item  & Interaction  & Density & FN \\
    \hline
    \multirow{2}{*}{synthetic dataset}
    &Movielens-100k & 943 & 1,682 & $100,000$ & $0.06304$ &17,057\\
    &Douban & 2,862 & 3,000 & $134,498$ & $0.01566$ &23,599 \\
    \hline
    \multirow{3}{*}{Real-world dataset}
    &Taobao & 22,976 & 29,149 & $439,305$ & $0.00066$ &-\\
    &Tmall & 10,000 & 14,965 & $461,899$ & $0.00309$ &- \\
    &Gowalla & 29,858 & 40,981 & $1,027,370$ & $0.00084$ &-\\
    \hline
    \end{tabular}
    \end{adjustbox}
  \end{center}
\end{table*}

\textbf{Datasets.} For simulation experiments, we generate synthetic datasets from two datasets Movielens-100k and Douban that have a relatively high user-item interaction density. Specifically, for a given user $u$ in Movielens-100k or Douban, we divide his interacted items into training set, test set and false negative set (denoted as $FN$ set) at a 4:1:1 ratio. The false negative set of user $u$ is denoted as $FN_{u}$. Instances in $FN_{u}$ are in fact positive items since they have interaction records with user $u$, but they are unobserved during recommender training. In simulation experiments, $FN$ set is split out to investigate the impact of false negatives on recommender training behavior.    

For real data experiments, three real-world datasets Taobao, Tmall and Gowalla, are utilized evaluate the performance of our method. All three datasets contain historical interactions among users and items only. Details are summarized in Table~\ref{tab:data}. For each dataset, we sort historical interactions \emph{w.r.t.} timestamps, then retain the latest $10\%$ records of each user to form test set. The rest of interactions are divided by 8:1 to obtain the training set and validation set.  Hyper-parameters are tuned according to model’s performance on validation set, and final results on test set are reported.
\\\\
\textbf{Evaluation metrics.} A rank list $R_u$ consisting of top $K$ items with highest socres will be provided to user $u$ for personalized recommendations. Recall@$K$ and NDCG@$K$ are leveraged as evaluation metrics as in most works, both of which assess the model’s capacity to find potentially relevant items, while NDCG@$K$ accounts more for the position of relevant items in $R_u$. According to \citet{shi2023theories}, a smaller value for $K$ necessitates a larger setting for $H$ to ensure a good performance. For easy comparison, we set $K=50$ as done in prior studies~\citep{chen2022learning,lian2020personalized,shi2023theories} and determine the optimal value for $H$ through experiments.
\\\\
\textbf{Baselines.} 
\label{baseline}
We conduct a comprehensive comparative analysis between the proposed PDNS method and various advanced negative sampling techniques employed in implicit collaborative filtering. The considered techniques are outlined as follows:

\begin{itemize}
\item \textbf{RNS}~\citep{DBLP:conf/uai/RendleFGS09}: Randomly selects unobserved user-item pairs as negatives.
\item
\textbf{DNS}~\citep{DBLP:conf/wsdm/RendleF14}: Adaptively selects the highest-scored item from the negative candidate pool.
\item 
\textbf{MixGCF}~\citep{DBLP:conf/kdd/HuangDDYFW021}: Specifically designed for GNN-based recommendation models, dynamically selects the highest-scored item using positive mixing and hop mixing techniques.
\item
\textbf{IRGAN}~\citep{DBLP:conf/sigir/WangYZGXWZZ17}: A GAN-based hard negative sampler that introduces reinforcement learning principles for model optimization.
\item
\textbf{AdvIR}~\citep{DBLP:conf/www/ParkC19}: Another GAN-based hard sampler that combines adversarial sampling and adversarial training to enhance information retrieval.
\item
\textbf{SRNS}~\cite{DBLP:conf/nips/DingQY0J20}: Mitigates false negatives by excluding items with low score variance in hard negative sampling.
\item
\textbf{GDNS}~\cite{zhu2022gain}: Prevents false negatives in hard negative sampling by favoring items with high expectational gain.
\item
\textbf{AdaSIR}~\citep{chen2022learning}: Utilizes importance resampling for hard negative sampling, with AdaSIR(U) and AdaSIR(P) respectively sampling candidate items from a uniform distribution and a popularity-based distribution.
\end{itemize}


\subsection{Experiments on synthetic datasets}
\label{sec:syn_experiment}
In simulation experiments, we aim to verify our finding that false negatives contribute to recommender’s over-fitting. The main idea is to see whether the over-fitting can be alleviated when avoiding a part of false negatives during negative sampling. 

The FN set generated from original dataset can be regards as a collection of false negatives from a ‘God’s-eye’ view. To control the impact of avoiding false negative instances in negative sampling, we manually disclose $c\times100\%$ of $FN$ set ($c\in[0,1]$, denoted as false coefficient) to recommender training process, and prevent selecting them as negative instances for model update. Note that $c=0$ represents a realistic case, training recommendation models with no knowledge about false negatives, and $c=1$ represents an ‘perfect’ situation that we avoid all false negatives we know (Experimental details are in Appendix~\ref{sec:supplement_syn_exp}).

\begin{figure}[t]
\centering
\subfigure[Movielens-100k, MF]{
\label{fig.synthetic_a}
\includegraphics[width=0.45\columnwidth]{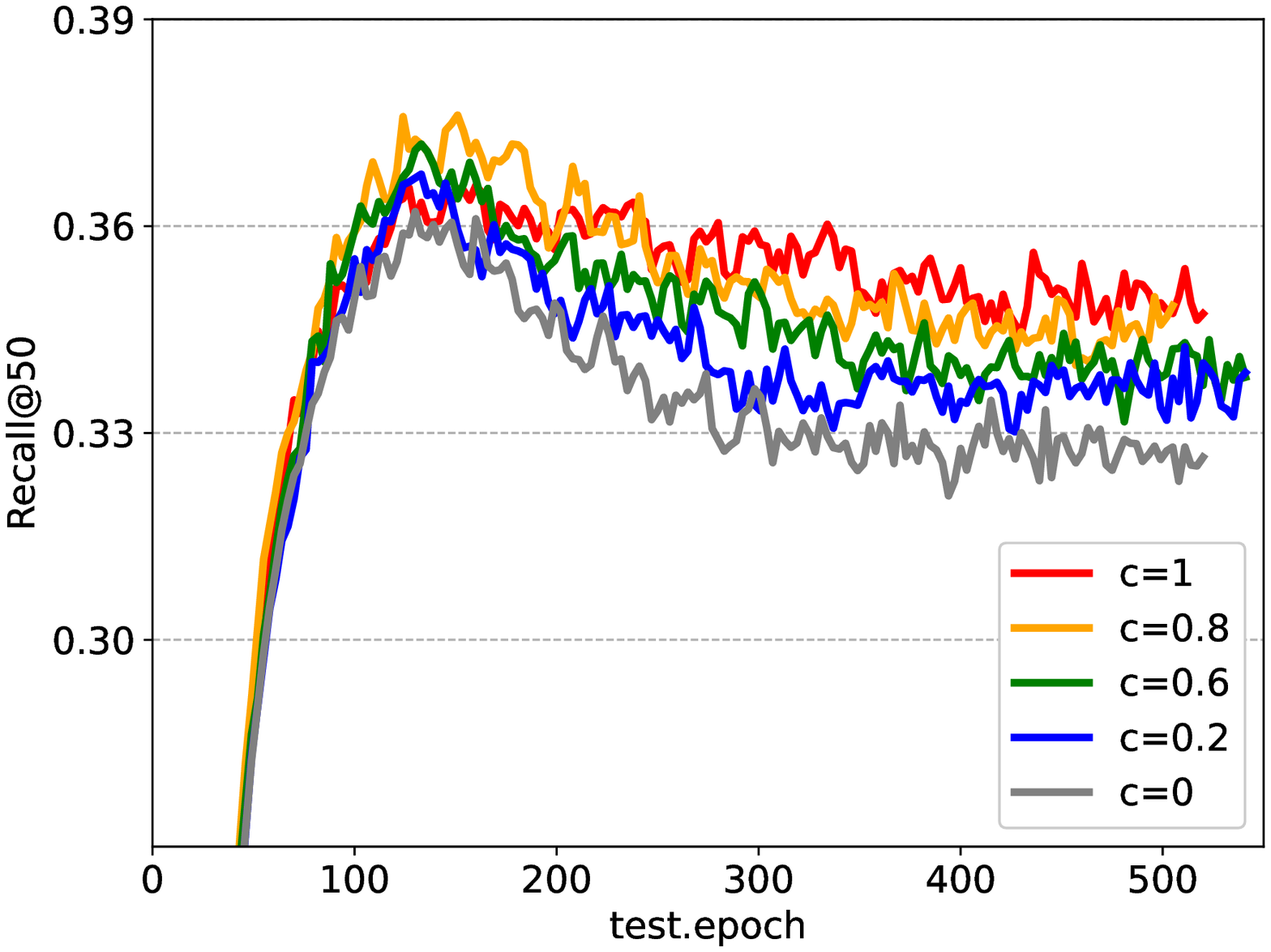}}
\subfigure[Douban, MF]{
\includegraphics[width=0.45\columnwidth]{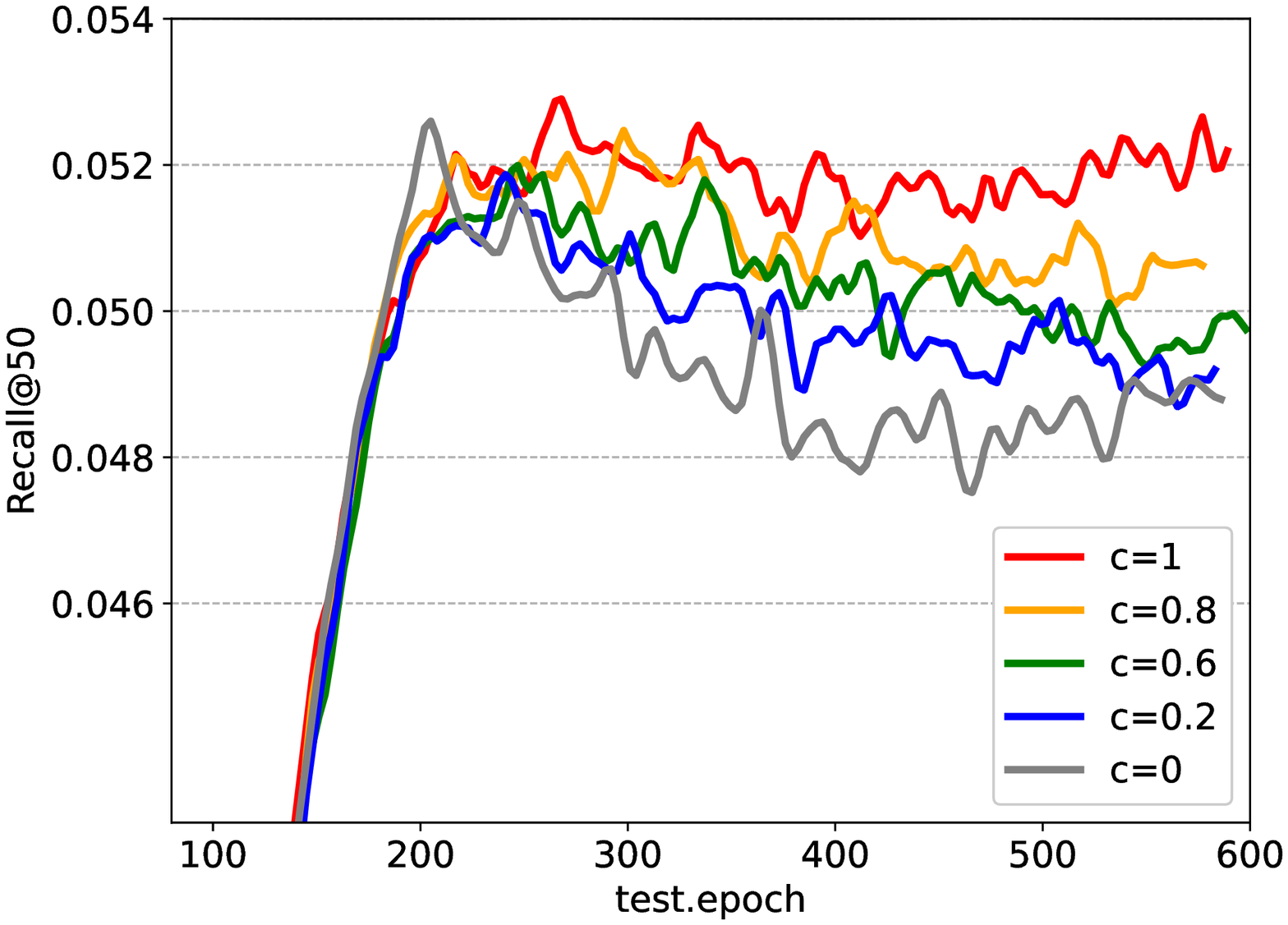}}
\subfigure[Movielens-100k, LightGCN]{
\includegraphics[width=0.45\columnwidth]{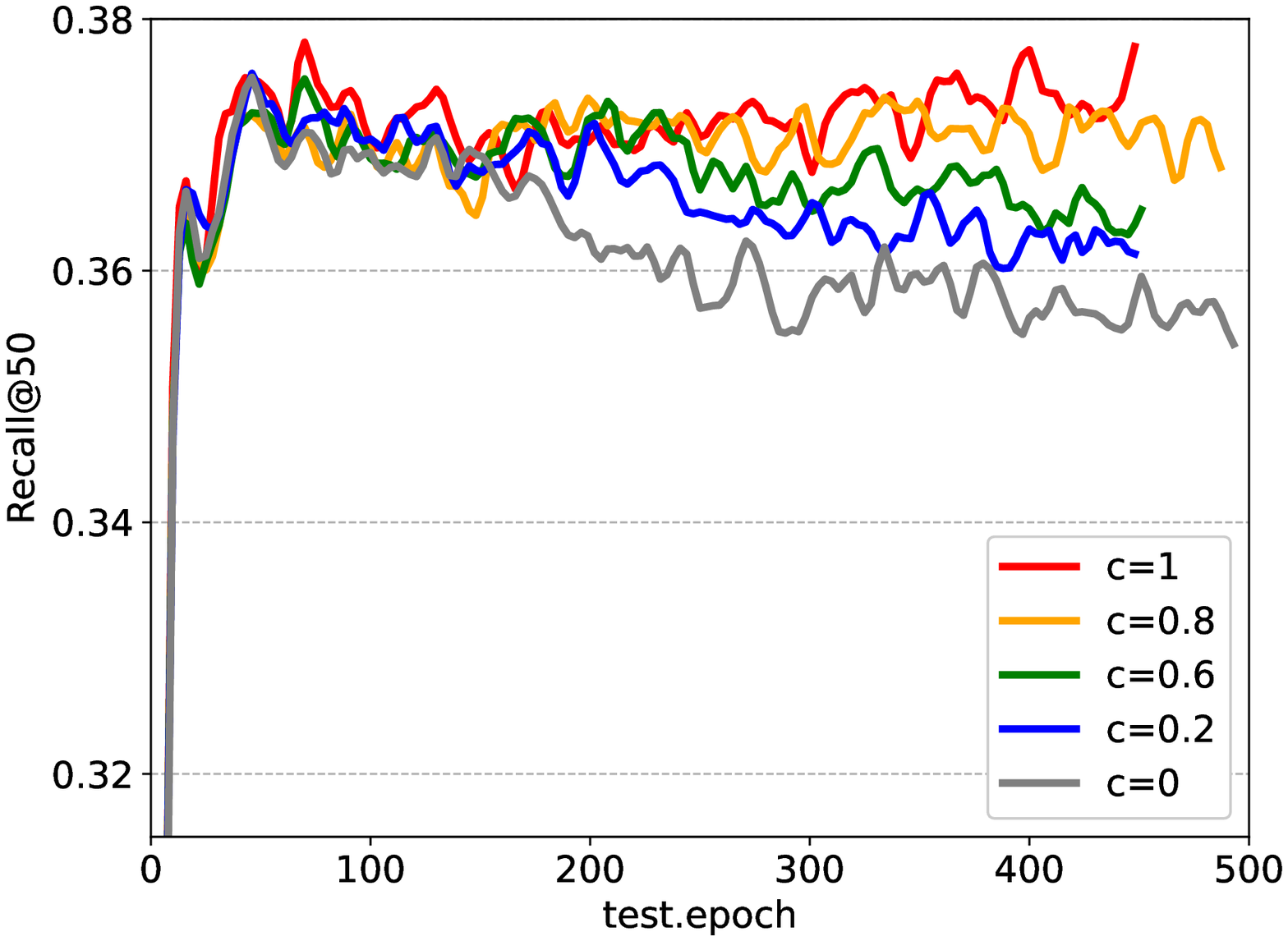}}
\subfigure[Douban, LightGCN]{
\includegraphics[width=0.45\columnwidth]{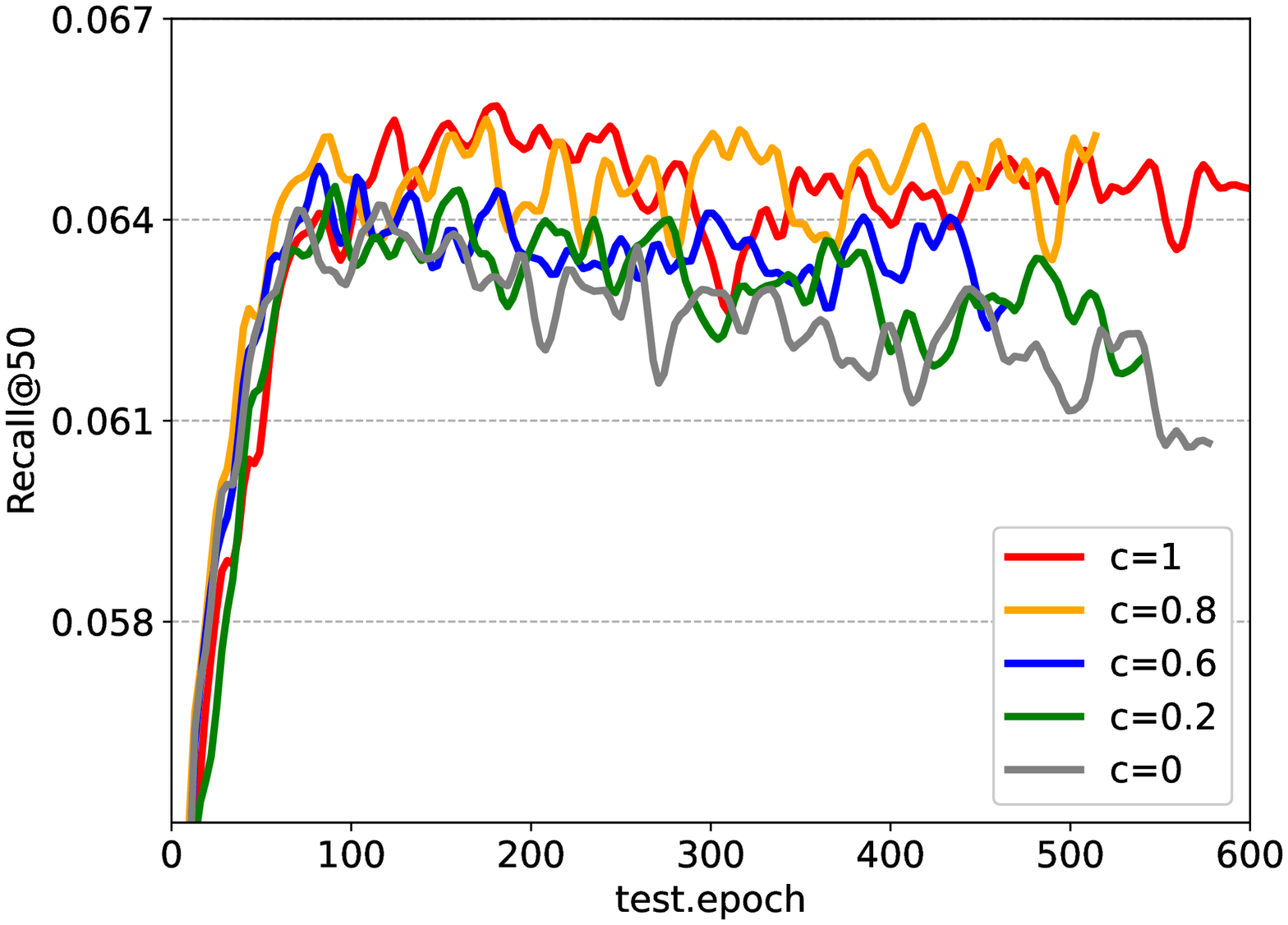}}
\caption{Average test results under different false coefficient $c$. With MF and LightGCN as recommendation models and DNS as negative sampling strategy.}
\label{fig:synthetic}
\end{figure}

With MF and LightGCN as two base models and DNS as the negative sampling strategy, experimental results on two synthetic datasets are shown in Figure~\ref{fig:synthetic} when false coefficient $c$ increases. We observe that when more false negatives are avoided during negative sampling, the over-fitting of the recommender is consistently mitigated on both synthetic datasets and under two base models. It verifies that incorrect selection of false negatives contributes to recommender’s over-fitting. Note that even when $c=1$, the over-fitting may still exists (particularly in Figure~\ref{fig.synthetic_a}). Its reason might be that our synthetic FN sets fail to contain all actual false negatives of original datasets. 

\subsection{Experiments on real-world datasets}
\label{sec:real_exp}

\begin{table*}[t]
  \begin{center}
    \caption{Performance comparison with LightGCN and MF as the base recommendation models. The last row is relative improvements of our method compared to the strongest baseline (underlined).}
    \label{tab:comparison}
    \begin{adjustbox}{width=\textwidth}
    \begin{tabular}{r|cc|cc|cc|cc|cc|cc}
    \toprule
          & \multicolumn{6}{c|}{\textbf{LightGCN}}        & \multicolumn{6}{c}{\textbf{MF}} \\
\cmidrule{2-13}          & \multicolumn{2}{c|}{Taobao} & \multicolumn{2}{c|}{Tmall} & \multicolumn{2}{c|}{Gowalla} & \multicolumn{2}{c|}{Taobao} & \multicolumn{2}{c|}{Tmall} & \multicolumn{2}{c}{Gowalla} \\
          & Recall & NDCG  & Recall & NDCG  & Recall & NDCG  & Recall & NDCG  & Recall & NDCG  & Recall & NDCG \\
    \midrule
    \multicolumn{1}{l|}{RNS} & 0.1041 & 0.0378 & 0.0895 & 0.0476 & 0.2712 & 0.1804 & 0.0811 & 0.0295 & 0.0692 & 0.0333 & 0.2397 & 0.1477 \\
    \multicolumn{1}{l|}{AdvIR} & 0.1052 & 0.0382 & 0.0917 & 0.0487 & 0.2710 & 0.1801 & 0.0911 & 0.0327 & 0.0792 & 0.0426 & 0.2603 & 0.1620 \\
    \multicolumn{1}{l|}{IRGAN} & 0.1049 & 0.0380 & 0.0904 & 0.0480 & 0.2799 & 0.1847 & 0.0859 & 0.0314 & 0.0771 & 0.0415 & 0.2421 & 0.1508 \\
    \multicolumn{1}{l|}{DNS} & 0.1107 & \underline{0.0410} & 0.0970 & 0.0504 & \underline{0.2848} & \underline{0.1878} & \underline{0.0999} & \underline{0.0360} & 0.0713 & 0.0382 & 0.2572 & 0.1608 \\
    \multicolumn{1}{l|}{SRNS} & 0.1086 & 0.0392 & 0.0967 & 0.0509 & 0.2819 & 0.1866 & 0.0968 & 0.0349 & 0.0801 & 0.0433 & 0.2588 & 0.1620 \\
    \multicolumn{1}{l|}{GDNS} & 0.1089 & 0.0393 & 0.0975 & 0.0516 & 0.2835 & 0.1865 & 0.0970 & 0.0352 & \underline{0.0816} & \underline{0.0440} & 0.2611 & 0.1638 \\
    \multicolumn{1}{l|}{MixGCF} & \underline{0.1110} & 0.0403 & 0.0963 & 0.0508 & 0.2840 & 0.1872 & -     & -     & - & - & -     & - \\
    \multicolumn{1}{l|}{AdaSIR(U)} & 0.1100 & 0.0401 & 0.1009 & 0.0530 & 0.2816 & 0.1857 & 0.0958 & 0.0336 & 0.0800 & 0.0429 & \underline{0.2711} & \underline{0.1712} \\
    \multicolumn{1}{l|}{AdaSIR(P)} & 0.1080 & 0.0389 & \underline{0.1016} & \underline{0.0538} & 0.2808 & 0.1853 & 0.0955 & 0.0333 & 0.0810 & 0.0438 & 0.2702 & 0.1700 \\
    \midrule
    \multicolumn{1}{l|}{PDNS} & \textbf{0.1162} & \textbf{0.0425} & \textbf{0.1051} & \textbf{0.0552} & \textbf{0.2957} & \textbf{0.1943} & \textbf{0.1030} & \textbf{0.0376} & \textbf{0.0872} & \textbf{0.0473} & \textbf{0.2803} & \textbf{0.1805} \\
          & 4.68\% & 3.66\% & 3.44\% & 2.60\% & 3.83\% & 3.46\% & 3.10\% & 4.44\% & 6.86\% & 9.18\% & 4.21\% & 5.90\% \\
    \bottomrule
    \end{tabular}%

    \end{adjustbox}
  \end{center}
\end{table*}

\textbf{Performance Comparison.}
\label{sec:per_com}
The comparison results between PDNS and baselines on three real-world datasets with LightGCN and MF as the base recommendation models are shown in Table~\ref{tab:comparison}, where the last row is the relative improvements of PDNS (in bold) compared to the strongest baseline (underlined). The observations are as follows:

\begin{figure*}
\centering
\subfigure[Taobao, LightGCN]{
\label{fig.taobao_lgn}
\includegraphics[width=0.22\textwidth]{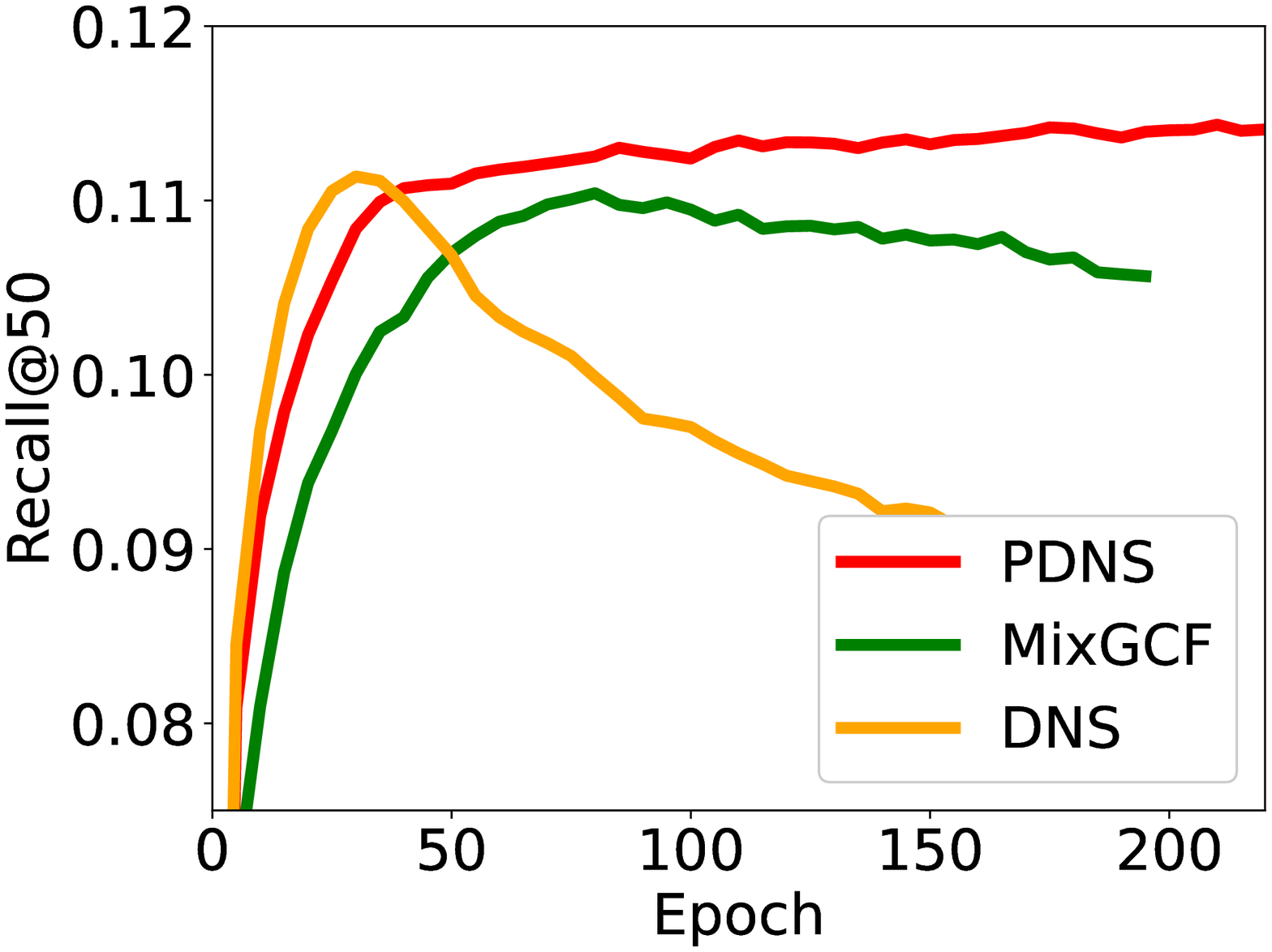}}
\subfigure[Tmall, LightGCN]{
\includegraphics[width=0.22\textwidth]{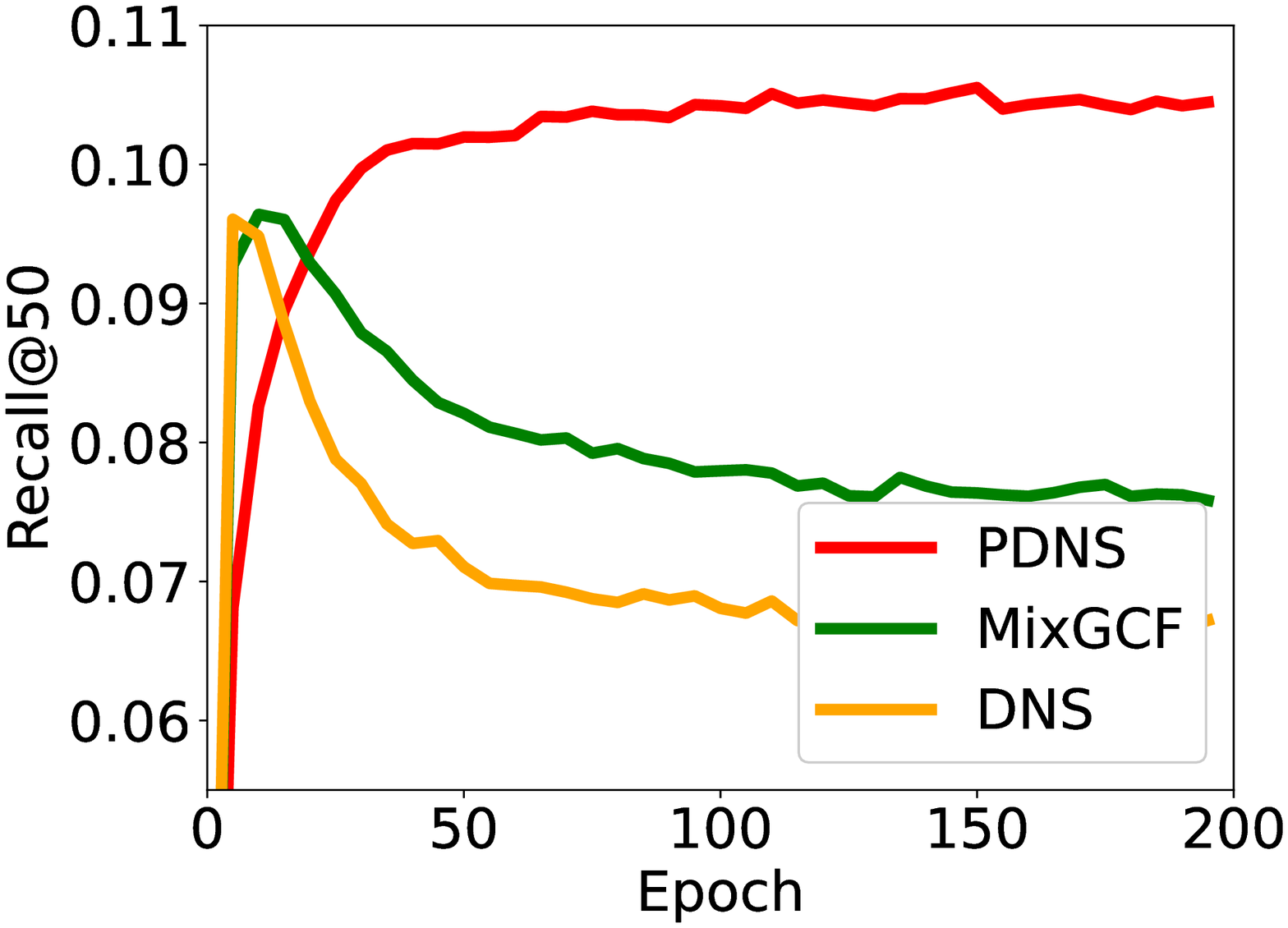}}
\subfigure[Gowalla, LightGCN]{
\includegraphics[width=0.22\textwidth]{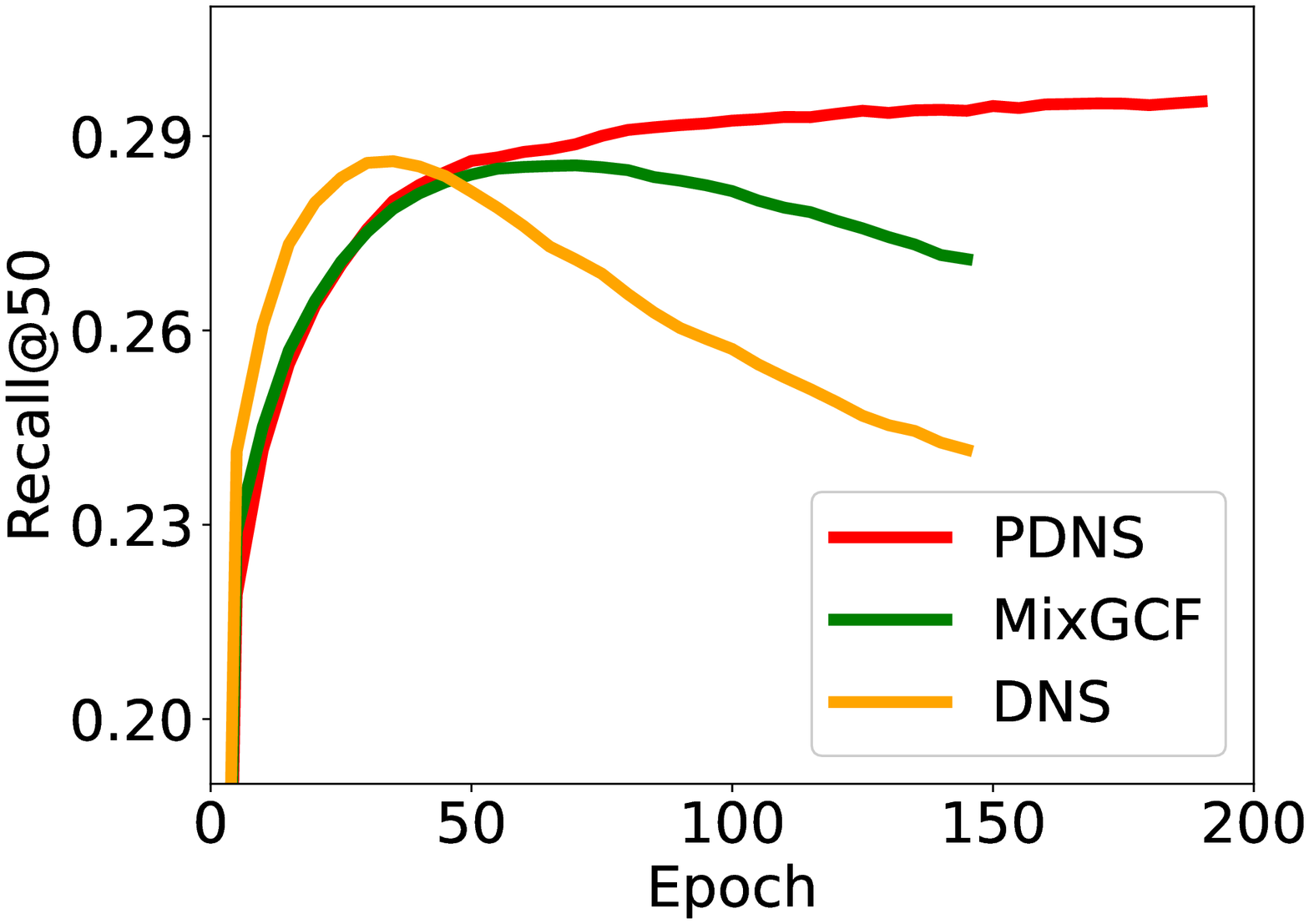}}


\subfigure[Taobao]{
\label{fig.H_taobao}
\includegraphics[width=0.22\textwidth]{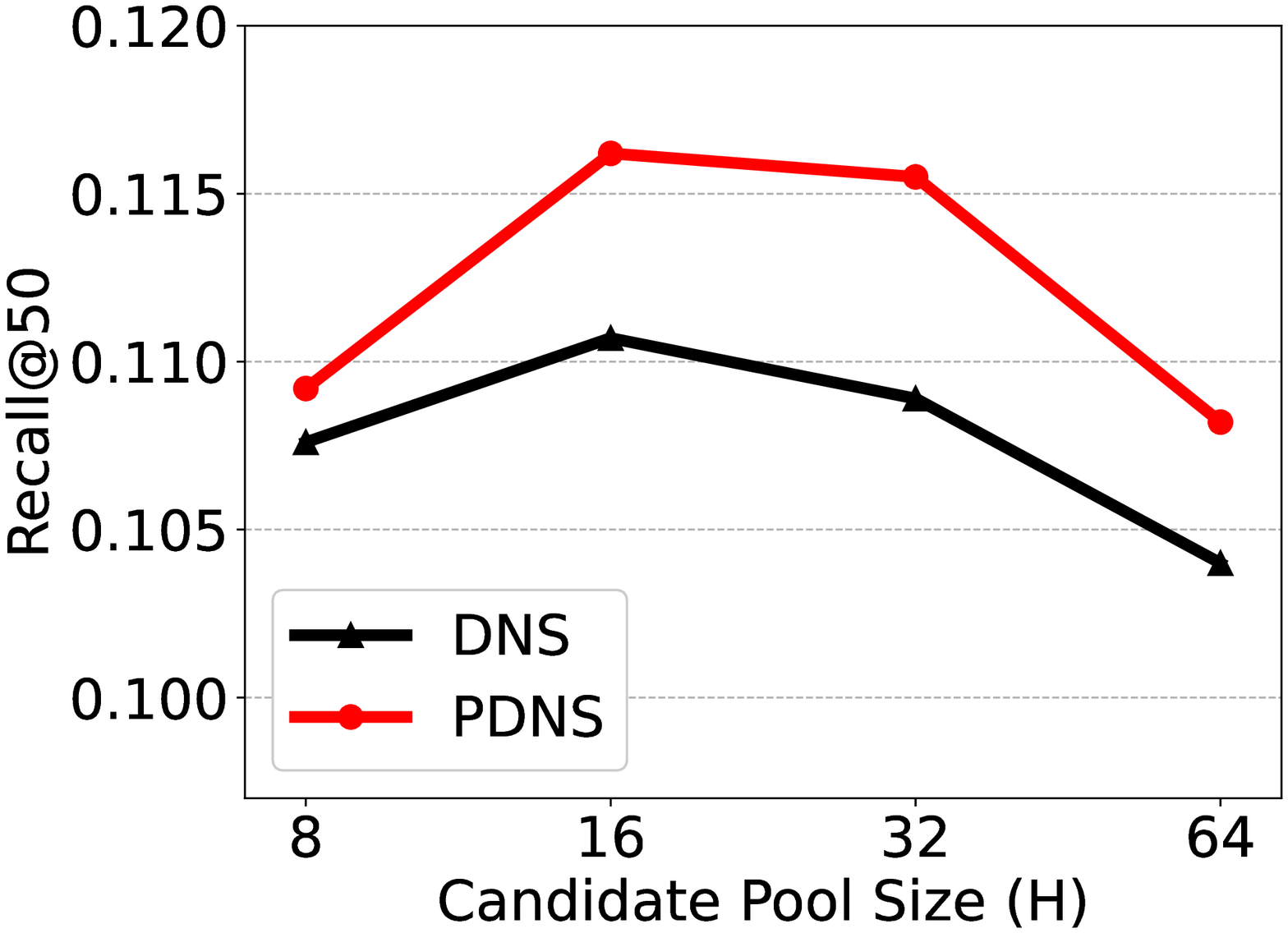}}
\subfigure[Tmall]{
\label{fig.H_tmall}
\includegraphics[width=0.22\textwidth]{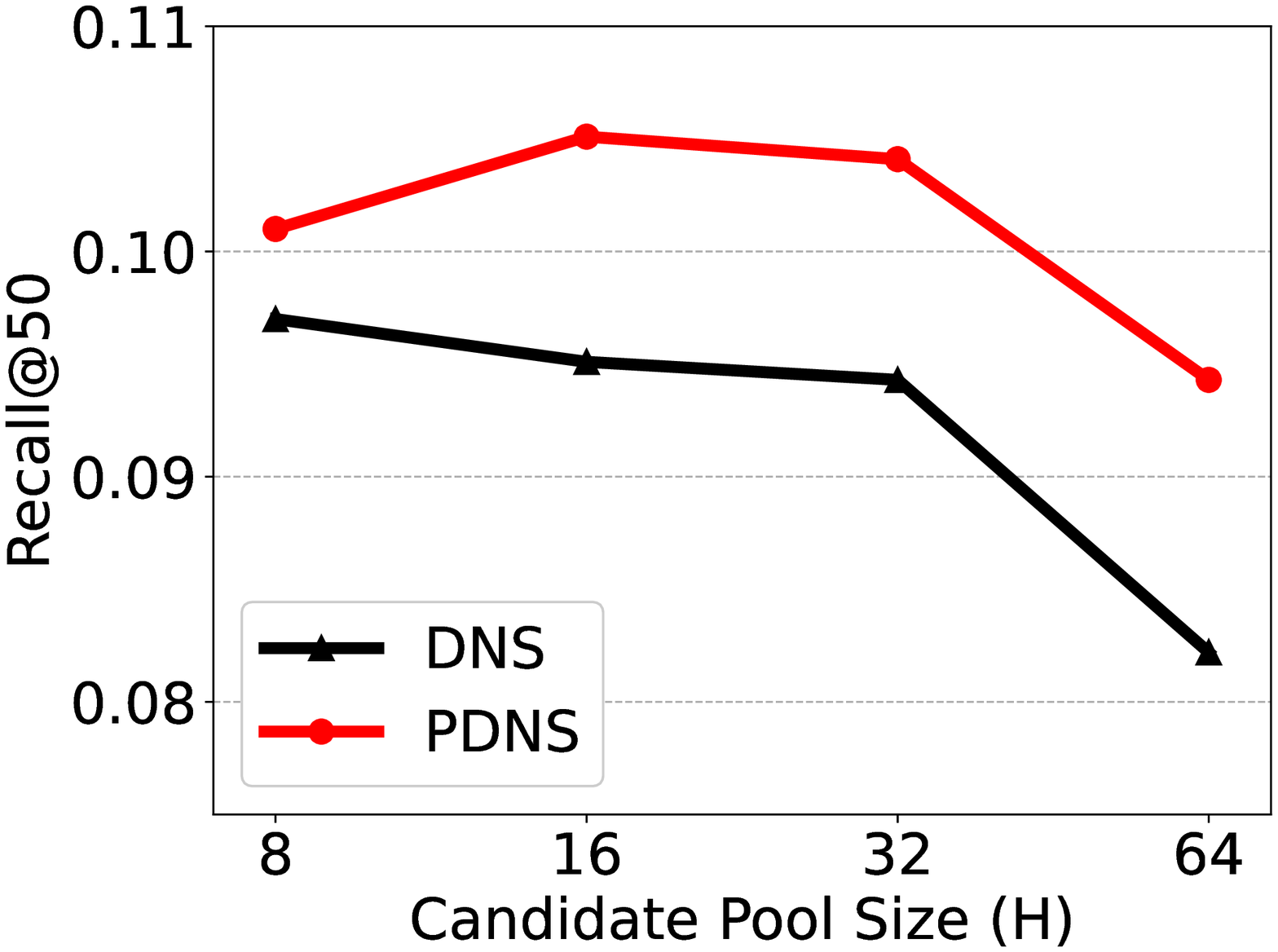}}
\subfigure[Gowalla]{
\label{fig.H_gowalla}
\includegraphics[width=0.22\textwidth]{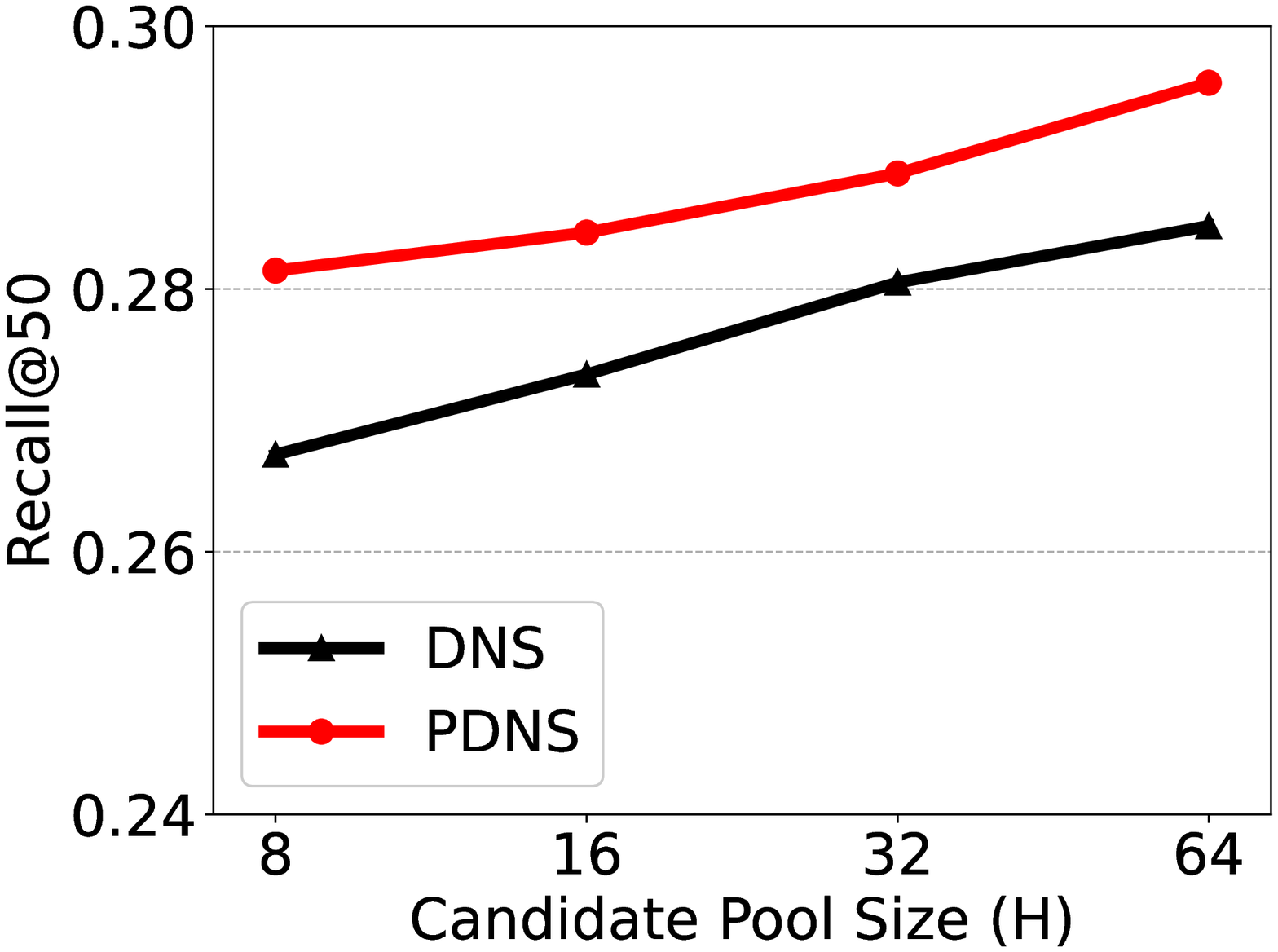}}

\caption{(a)-(c): Training curves on test sets \emph{w.r.t.} PDNS, DNS and MixGCF. (d)-(f): Recall@50 vs. different candidate pool size $H$, comparing PDNS with DNS under LightGCN model.}
\label{fig:robust_H}
\end{figure*}

\begin{itemize}
    \item Under two different base recommendation models, PDNS consistently outperforms all baselines across three datasets, the relative improvements compared to the second best reach $3.10\%\sim6.86\%$ on test Recall@50. 
    \item PDNS can not only suit GNN-based recommendation models like LightGCN, but also can be plugged into other types of methods such as MF, because the idea, preventing few hardest negatives dominating model updates, is general and model-independent.
    \item LightGCN, as a state-of-the-art GNN-based method, yields better performance than MF across all datasets. PDNS demonstrates an average relative improvement of $3.98\%$ across three datasets compared to the strongest baseline, when adopting LightGCN as the base model. This improvement is slightly less than the $4.72\%$ observed when MF serves as the base model. This suggests that PDNS has the capability to release the predictive potential of simple models. 
\end{itemize}
\textbf{Robustness analysis.} In this subsection, we exhibit the robustness of the proposed method. We mainly compare PDNS with two state-of-the-art hard negative sampling strategies DNS and MixGCF, where MixGCF also adopts positive mixing technique, but the mixing coefficient $\alpha$ is sampled from uniform distribution $\mathbf{U}(0,1)$. The training curves of PDNS, DNS and MixGCF on three test sets under LightGCN are drawn in Figure~\ref{fig:robust_H}. As can be seen from Figure~\ref{fig:robust_H} (a)-(c), the training curves under LightGCN share the same pattern that DNS exhibits the most severe over-fitting, while MixGCF is less but still notably severe. PDNS, however, is not only more effective, but also more stable on all datasets. To summarize, adopting PDNS in recommender training is more resilience to over-fitting, demonstrating better robustness of PDNS towards potential false negative samples in recommendation tasks discussed in Section~\ref{sec:over-fitting}.
\begin{table}[t]
  \begin{center}
    \caption{Performance of PDNS with different candidate pool size $H$.}
    \label{tab:H_study}
    \begin{adjustbox}{width=\columnwidth}
        \begin{tabular}{ll|cc|cc|cc}
        \hline && \multicolumn{2}{c|}{ Taobao } & \multicolumn{2}{c|}{ Tmall } & \multicolumn{2}{c}{ Gowalla } \\
         && Recall & NDCG & Recall & NDCG & Recall & NDCG \\
        \hline
        \hline
        \multirow{4}{*}{LightGCN+PDNS}
        &$H=8$   & 0.1092 & 0.0412 & 0.1010 & 0.0529 & 0.2814 & 0.1806 \\
        &$H=16$  & \underline{0.1162} & \underline{0.0425} & \underline{0.1051} & \underline{0.0552} & 0.2843 & 0.1833 \\
        &$H=32$  & 0.1155 & 0.0424 & 0.1041 & 0.0550 & 0.2888 & 0.1901 \\
        &$H=64$  & 0.1082 & 0.0408 & 0.0943 & 0.0517 & \underline{0.2957} & \underline{0.1943} \\
        \hline
        \hline
        \multirow{4}{*}{MF+PDNS}
        &$H=8$   & \underline{0.1053} & \underline{0.0383} & 0.0874 & 0.0453 & 0.2494 & 0.1585 \\
        &$H=16$  & 0.0895 & 0.0330 & 0.0890 & 0.0481 & 0.2677 & 0.1716 \\
        &$H=32$  & 0.0596 & 0.0222 & \underline{0.0896} & \underline{0.0483} & 0.2800 & 0.1784 \\
        &$H=64$  & 0.0411 & 0.0154 & 0.0525 & 0.0274 & \underline{0.2808} & \underline{0.1803} \\
        \hline
        \end{tabular}
    \end{adjustbox}
  \end{center}
\end{table}
\\\\
\textbf{Hardness study.} As shown in Table~\ref{tab:H_study}, we conduct hardness study of PDNS on three real-world datasets under LightGCN and MF by changing the hardness level $H$ (the size of negative candidate pool), where the best results are underlined. Recall that the larger the $H$, the harder the negative sampled. In addition, we draw the experimental results of PDNS on three datasets under LightGCN at the last row of Figure~\ref{fig:robust_H}, and compare it with DNS. We observe that:

\begin{itemize}
    \item For Taobao and Tmall, both DNS and PDNS manifest the same trend with the increase of $H$, that is, reaching the peak and then suffering from degradation. Such a situation is consistent with our discussion about false negatives in Section~\ref{sec:over-fitting}, \emph{i.e.}, harder negatives are more likely to be the false negatives, which will confuse the recommender during training procedure. As for Gowalla, due to larger user and item sets, harder negatives ($H=64$) are preferred for model optimization.
    \item PDNS outperforms DNS by a large margin under different $H$ settings, indicating the better capacity of PDNS to offer valuable information from hard negatives for model updating.\\
\end{itemize}
\textbf{Soft factor study.} In Section~\ref{equi_alg}, we derive soft BPR loss in equivalent algorithm of PDNS. The soft factor $\beta$ is critical for performance gains, which is empirically set as a small number to balance the gradients of negative samples. In Table~\ref{tab:study_b}, we demonstrate the performance of PDNS when varying $\beta$ in $\{0.01, 0.05, 0.1, 0.2, 0.3, 0.4\}$ under LightGCN and MF. The best results are underlined. 
\begin{table}[th]
  \begin{center}
    \caption{Performance of PDNS with different soft factor $\beta$.}
    \label{tab:study_b}
    \begin{adjustbox}{width=\columnwidth}
        \begin{tabular}{ll|cc|cc|cc}
        \hline && \multicolumn{2}{c|}{ Taobao } & \multicolumn{2}{c|}{ Tmall } & \multicolumn{2}{c}{ Gowalla } \\
         && Recall & NDCG & Recall & NDCG & Recall & NDCG \\
        \hline
        \hline
        \multirow{6}{*}{LightGCN+PDNS}
        &$\beta=0.01$  & 0.0545 & 0.0192 & 0.0603 & 0.0271 & \underline{0.2957} & \underline{0.1943} \\
        &$\beta=0.05$  & 0.1123 & 0.0410 & 0.0430 & 0.0232 & 0.2872 & 0.1898 \\
        &$\beta=0.1$   & \underline{0.1162} & \underline{0.0425} & \underline{0.1051} & \underline{0.0552} & 0.2859 & 0.1884 \\
        &$\beta=0.2$   & 0.1152 & 0.0421 & 0.1009 & 0.0536 & 0.2843 & 0.1868 \\
        &$\beta=0.3$   & 0.1115 & 0.0417 & 0.0995 & 0.0531 & 0.2823 & 0.1859 \\
        &$\beta=0.4$   & 0.1137 & 0.0420 & 0.0978 & 0.0523 & 0.2823 & 0.1854 \\
        \hline
        \hline
        \multirow{6}{*}{MF+PDNS}
        &$\beta=0.01$  & 0.0018 & 0.0004 & 0.0603 & 0.0233 & 0.0024 & 0.0010 \\
        &$\beta=0.05$  & 0.0318 & 0.0139 & 0.0103 & 0.0045 & 0.0978 & 0.0624 \\
        &$\beta=0.1$   & \underline{0.1053} & \underline{0.0383} & 0.0316 & 0.0144 & 0.2178 & 0.1376 \\
        &$\beta=0.2$   & 0.0954 & 0.0348 & 0.0759 & 0.0419 & \underline{0.2808} & \underline{0.1803} \\
        &$\beta=0.3$   & 0.0879 & 0.0316 & \underline{0.0896} & \underline{0.0483} & 0.2764 & 0.1779 \\
        &$\beta=0.4$   & 0.0816 & 0.0287 & 0.0850 & 0.0464 & 0.2693 & 0.1732 \\
        \hline
        \end{tabular}
    \end{adjustbox}
  \end{center}
\end{table}
We observe that for different recommendation models and datasets, the $\beta$ need to be tuned properly to achieve the best performance. In some cases, when $\beta$ set too small, the recommender may fail to learn from data. For example, the test Recall@50 on Taobao is only 0.0545 when $\beta=0.01$ when employing LightGCN as the base model (the best is 0.1162). Although we need to choose a proper $\beta$, PDNS enables effortless tuning and easy implementation owing to its small hyper-parameter search space.

\section{Related work}
In order to obtain high-quality negatives in implicit CF, several recent efforts work on hard negative samplers via emphasizing items with large matching scores~\citep{DBLP:conf/nips/DingQY0J20, DBLP:conf/wsdm/RendleF14, DBLP:conf/kdd/HuangSSXZPPOY20}. For example, IRGAN~\citep{DBLP:conf/sigir/WangYZGXWZZ17} and AdvIR~\citep{DBLP:conf/www/ParkC19} leverage generative adversarial networks to generate hard negative items. Adaptive hard negative sampling techniques such as DNS~\citep{DBLP:conf/wsdm/RendleF14}, SRNS~\citep{DBLP:conf/nips/DingQY0J20}, MixGCF~\citep{DBLP:conf/kdd/HuangDDYFW021} and GDNS~\cite{zhu2022gain} emerge as the advanced sampling strategies in implicit CF. AdaSIR~\cite{chen2022learning} and Softmax-v($\rho, N$)~\cite{shi2023theories} sample high-quality negative instances from an approximated softmax distribution. Furthermore, \citet{shi2023theories} has recently advanced the theory of hard negative sampling, demonstrating that optimizing BPR under DNS serves as a precise estimator of One-way Partial AUC. This study establishes a linkage between Top-$K$ metrics and the utilization of hard negative sampling.

There are a limited number of works in addressing the false negative problem within the context of hard negative sampling. SRNS first observes that both hard negatives and false negatives are assigned to high scores, yet false negatives have lower score variances comparatively. Subsequently, \citet{cai2022hard} introduces a Coupled Estimation Technique to correct the bias caused by false negative instances, and \citet{zhu2022gain} utilizes expectational gain gap to guide false negative filtering. 

Developed from $mixup$ technique~\citep{ DBLP:conf/iclr/ZhangCDL18}, MoCHi first injects positive information into hard negative embeddings to derivate even harder negatives~\citep{DBLP:conf/nips/KalantidisSPWL20}. Subsequently, MixGCF adopts the positive mixing technique in MoCHi to synthesizes negatives for implicit CF task~\citep{DBLP:conf/kdd/HuangDDYFW021}. In this paper, we deeply investigate why positive information is critical in hard negative synthesizing.

\section{Conclusion}

In this paper, we comprehensively study the over-fitting phenomenon in hard negative sampling of implicit collaborative filtering, and empirically establish the connection between the over-fitting and the unintentional selection of false negatives. Subsequently, to address the over-fitting problem, we conduct study on the positive mixing technique. In previous works, a critical coefficient in positive mixing is intuitively set sampled from a uniform distribution (\emph{e.g.} $\mathbf{U}(0,1)$ is employed by~\citet{DBLP:conf/kdd/HuangDDYFW021} and $\mathbf{U}(0,0.5)$ by~\citet{DBLP:conf/nips/KalantidisSPWL20}). Our empirical findings reveal that fixing the critical coefficient to a constant value, tuned within the range of $(0.7,1)$, can yield great performance in recommendation pairwise learning and mitigate the effect of false negatives.  

Inspired by aforementioned study, we introduce a positive dominated negative synthesizing strategy PDNS. PDNS is a general hard mining technique which can be easily plugged into various recommendation models to improve personalized recommendation accuracy, and it can significantly alleviate over-fitting caused by incorrect selection of false negatives in hard negative sampling process. Extensive experiments on three datasets demonstrate the advantages of PDNS in terms of both effectiveness and robustness. In future work, we will continue to enhance recommender systems by improving model’s capacity to identify and avoid false negatives.

\appendix
\section{appendix}
\subsection{Running environment}
We conduct experiments on a single Linux server equipped with an AMD EPYC 7543, 128GB RAM and NVIDIA GeForce RTX 4090. We implement PDNS in Python 3.9 and PyTorch 1.9.1.

\subsection{Dataset details}
Five datasets used in simulation experiments and real data experiments are listed below:
\begin{itemize}
    \item {Movielens-100k\footnote{\url{https://grouplens.org/datasets/movielens/100k/}}} is a movie-rating dataset which contains ratings varying from 1 to 5. In the implicit feedback setting, we treat all these rating records as positive instances.
    \item {Douban\footnote{\url{https://www.douban.com}}} includes various ratings and reviews towards music, books and movies, derived from a social platform douban. We regard all ratings and reviews as positive examples, and use the pre-processed data in the work GCMC\citep{ DBLP:journals/corr/BergKW17}. 
    \item {Taobao\footnote{\url{https://tianchi.aliyun.com/dataset/dataDetail?dataId=649}}} is derived from e-commerce platform taobao, containing various user behaviors, \emph{e.g.}, click, add to cart and buy. To mitigate data sparsity, we regard these different interaction behaviors as positive labels for prediction task. For dataset quality, we randomly choose a subset of users from those with at least 10 interaction records, a.k.a. 10-core setting.
    \item {Tmall\footnote{\url{https://tianchi.aliyun.com/dataset/dataDetail?dataId=121045}}} is collected from another e-commerce platform tmall. This dataset covers a special period of a large sale event, thus the trading volume is relatively larger than normal.  Similarly, we treat multiple behaviors as positive and adopt 10-core setting for customer filtering.
    \item {Gowalla} is used for recommendation tasks in social networks~\citep{liang2016modeling,DBLP:conf/sigir/Wang0WFC19}, consisting of check-in records with users’ locations. We directly utilize the pre-processed dataset in \citet{DBLP:conf/sigir/Wang0WFC19} for our recommender evaluation. 
\end{itemize}


    

\subsection{Parameter tunning}

\begin{table}[!h]
  \begin{center}
    \caption{Optimal hyperparameters of PDNS on synthetic datasets.}
    \label{tab:opthyp_syn}
        \begin{tabular}{ll|cccccc}
        \hline & & $H$ & $lr$ &  $F$ & $L$ & $\lambda$ \\
        \hline \multirow{2}{*}{ MF }  
        &Movielens-100k & 32  & 0.001 & 32  & 1 & 0.01 \\
        &Douban & 32    & 0.002 & 32   & 1   & 0.01 \\
        \hline \multirow{2}{*}{ LightGCN }
        &Movielens-100k & 32   & 0.005 & 32  &  -   & 0.01 \\
        &Douban & 32   & 0.002 & 32    &  -    & 0.01 \\ 
        \hline
        \end{tabular}
  \end{center}
\end{table}

\begin{table}[!h]
  \begin{center}
    \caption{Optimal hyperparameters of PDNS on real-world datasets.}
    \label{tab:opthyp_real}
        \begin{tabular}{ll|ccccccc}
        \hline & & $H$ & $lr$ &  $F$ & $L$ & $\lambda$ & $\beta$\\
        \hline \multirow{3}{*}{ LightGCN }  &Taobao & 16    & 0.001 &  64    & 3     & 0.001 & 0.1 \\
        &Tmall & 16    & 0.005 &  64    & 3     & 0.001 & 0.1 \\
        &Gowalla & 64    & 0.005 &  64    & 3     & 0.0001 & 0.01 \\

        \hline \multirow{3}{*}{ MF}    &Taobao & 8     & 0.001 &  32    &  -     & 0.01  & 0.1 \\
        &Tmall & 32    & 0.001 &  32    & -       & 0.01  & 0.3 \\
        &Gowalla & 64    & 0.001 &  64    & -      & 0.01  & 0.2 \\
        \hline
        \end{tabular}
  \end{center}
\end{table}

We conduct a grid search to find the optimal hyperparameters. The optimizer is Adam and batch size $B$ is set as 2048 for all experiments. As for recommenders, we fix the number of aggregation layers $L$ in LightGCN as 3 by default. Embedding dimension $F$ is tuned in $\left\{8,16,32,64\right\}$. Candidate pool size $H$ of DNS, PDNS, and MixGCF is searched in $\left\{8,16,32,64\right\}$ and positive coefficient $\beta$ is tuned in $\left\{0.01,0.05,0.1,0.2,0.3,0.4\right\}$. Besides, we search learning rate $lr$ and regularization term $\lambda$ in $\left\{0.0005,0.001,0.002,0.005,0.01\right\}$ and $\left\{0,0.1,0.01,0.001,0.0001\right\}$ respectively. For simulation experiments, we tune hyper-parameters according to the average best Recall@50 on the training set. For real data experiments, we select the optimal hyper-parameters according to the best Recall@50 on the validation set, and report experimental results with the best validation Recall@50. We repeat five times with different random seeds and calculate the average for each experiment setting. Detailed information is listed in Table~\ref{tab:opthyp_syn} and Table~\ref{tab:opthyp_real}.For training curve of PDNS on Taobao under LightGCN in Figure~\ref{fig.taobao_lgn}, the learning rate $lr$ is set as 0.005 before epoch 40 to speed up model training.

For baselines mentioned in this work, to achieve fair comparison with PDNS in real data experiments, their hyper-parameters are also finely tuned guided by the best validation Recall@50. The hyper-parameter search space of DNS and MixGCF is the same with PDNS, and a pretrained model under RNS is used to initialize two GAN-based hard negative samplers. 

\subsection{Experiments on synthetic datasets}
\label{sec:supplement_syn_exp}
In Section~\ref{sec:syn_experiment}, to investigate whether avoiding false negatives in training process can mitigate recommender’s over-fitting, we manually reveal and avoid $c\times100\%$ of FN set in negative sampling with false coefficient $c$ in $[0,0.2,0.4,0.6,0.8]$. This progress is conducted segmentally. Specifically, we divide $c\times100\%$ of FN into several subsets, with each subset containing $20\%$ false negative instances in FN. Every 50 epochs after the validation Recall@50 reach the highest value, we disclose one subset to the recommender training, and prevent selecting all false negatives we know to update the model. 

In experiments, the points disclosing the first subset of false negative are: epoch 150 for Movielens100k-MF, epoch 220 for Douban-MF, epoch 100 for Movielens100k-LightGCN and epoch 100 for Douban-LightGCN.

\clearpage
\bibliographystyle{ACM-Reference-Format}
\bibliography{PDNS}

\end{document}